\newcommand\ha{{H$\alpha$}}
\newcommand\hb{{H$\beta$}}

\newcommand\kms{\:\rm{km\,s^{-1}}}

\newcommand\oiL{[\ion{O}{1}] $\lambda 6300$}

\newcommand\oiiiL{[\ion{O}{3}] $\lambda 5007$}
\newcommand\SiiLL{[\ion{S}{2}] $\lambda\lambda 6716, 6731$}
\newcommand\siiL{[\ion{S}{2}] $\lambda 6725$}
\newcommand\NiiLL{[\ion{N}{2}] $\lambda\lambda 6548, 6583$}

\newcommand\nicL{[\ion{Ni}{2}] $\lambda 7378$}

\newcommand\FeiiL{[\ion{Fe}{2}] 1.644 $\mu$m}

\newcommand\arii{[\ion{Ar}{2}]}

\newcommand\sii{[\ion{S}{2}]}
\newcommand\siii{[\ion{S}{3}]}

\newcommand\oi{[\ion{O}{1}]}
\newcommand\oii{[\ion{O}{2}]}
\newcommand\oiii{[\ion{O}{3}]}
\newcommand\feii{[\ion{Fe}{2}]}
\newcommand\nic{[\ion{Ni}{2}]}

 \newcommand{\MSOL}{\mbox{$\:M_{\sun}$}}

\documentclass[modern]{aastex631}
\usepackage[utf8]{inputenc}
\usepackage{natbib}
\usepackage{comment}
\usepackage{graphicx}
\usepackage{comment}
\usepackage{amssymb}
\usepackage{hyperref}
\turnoffediting

\shorttitle{HST/WFC3 Imaging of the Crab Nebula}
\shortauthors{Blair et al.}

\begin{document}

\title{The Crab Nebula Revisited Using HST/WFC3}

\correspondingauthor{W. P. Blair}
\email{wblair@jhu.edu}

\author[0000-0003-2379-6518]{William P. Blair}
\affil{The William H. Miller III Department of Physics and Astronomy, 
Johns Hopkins University, 3400 N. Charles Street, Baltimore, MD, 21218; wblair@jhu.edu}

\author[0000-0001-8858-1943]{Ravi Sankrit}
\affiliation{Space Telescope Science Institute,
3700 San Martin Dr., Baltimore MD 21218, USA;
rsankrit@stsci.edu}

\author[0000-0002-0763-3885]{Dan Milisavljevic}
\affiliation{Purdue University, Department of Physics and Astronomy, 525 Northwestern Ave, West Lafayette, IN 47907, USA}
\affiliation{Integrative Data Science Initiative, Purdue University, West Lafayette, IN 47907, USA; dmilisav@purdue.edu}

\author[0000-0001-7380-3144]{Tea Temim}
\affiliation{Department of Astrophysical Sciences, Princeton University, 4 Ivy Ln, Princeton, NJ 08544, USA; temim@astro.princeton.edu}

\author[0000-0002-3362-7040]{J. Martin Laming}
\affiliation{Space Science Division, Code 7684, Naval Research Laboratory, Washington DC 20375, USA; j.m.laming.civ@us.navy.mil}

\author[0000-0002-6986-6756]{Patrick Slane}
\affiliation{Center for Astrophysics $\vert$ Harvard \& Smithsonian, 60 Garden Street, Cambridge, MA 02138, USA; pslane@cfa.harvard.edu}

\author[0009-0003-0122-9472]{Ziwei Ding}
\affiliation{Purdue University, Department of Physics and Astronomy, 525 Northwestern Ave, West Lafayette, IN 47907, USA;  ding329@purdue.edu}

\author[0000-0002-3074-9608]{Thomas Martin}
\affiliation{D\'epartement de physique, de g\'enie physique et d’optique, Universit\'e Laval, 1045 avenue de la m\'edecine, Qu\'ebec, QC G1V 0A6, Canada}
\affiliation{C\'egep Garneau, 1660 Boulevard de l’Entente, Qu\'ebec, QC G1S 4S3, Canada; tmartin@cegepgarneau.ca}

\begin{abstract}

It has been over 24 years since the iconic Crab Nebula has been visited by the high spatial resolution eye of the Hubble Space Telescope. The expanding nebula is dynamic on these timescales, with many of the outer filaments of the nebula known to show proper motions of 0.3\arcsec\ or more per year.  Over time, it has become increasingly difficult to compare the fine scale structure of the nebula with recent data at other wavelengths.  We have re-observed the Crab in an HST Cycle 31 program using the WFC3 camera and filters similar to those previously used to make the existing mosaic that dates from 1999-2000 and was obtained with the WFPC2 camera.  Two central fields were observed with the F487N filter, providing an uncontaminated hydrogen band for comparison.
We also observed two primarily continuum band filters (F547M and F763M), allowing us to study the optical synchrotron nebula component of the Crab's emission.  We compare these new data to the first epoch of WFPC2 data as well as to more contemporaneous NIR/MIR imagery from JWST.  Finally, we highlight two previously unrecognized groupings of filaments with similar emission characteristics that are nearly diametrically opposed from the pulsar but whose origin remains uncertain.


\end{abstract}



\section{Introduction}

The Crab Nebula is the iconic and well-studied remnant of SN 1054 CE \citep{trimble77}.  It is thought to have arisen from a precursor star of 8-10 \MSOL, near the lower limit for core collapse supernovae (SNe).  Its known age, proximity (d = 2.0 kpc\footnote{We note the recent paper by \citet{fraser19}, who used Gaia DR2 data to argue for a significantly larger distance near 3.4 kpc. However, this result has not held up to scrutiny, and we adopt the canonical value of 2 kpc, consistent with many previous papers.}) and only moderate extinction ($E(B - V)$ = 0.3 - 0.5 mag) have made it a prime object of study across the electromagnetic spectrum and a key touchstone for our understanding of SNe and young SN remnants (see review articles by \citet{davidson85} and \citet{hester08}).

Early HST studies of the Crab using WFPC2 concentrated on the synchrotron nebular structure \citep{hester95} and characterizing the filamentary structures and ionization characteristics of the expanding debris field \citep{hester96, blair97, sankrit98}.  However, all of these studies involved a single WFPC2 field that covered the pulsar and central bright filaments, primarily to the north and west of the pulsar.  It was not until 1999-2000 that HST program 8222 obtained full coverage of the nebular and synchrotron structure of the Crab in three optical emission lines (\oiL, \oiiiL, and \SiiLL) plus a continuum-dominated band centered near 5500 \AA.  These data were used to make the famous HST color mosaic of the Crab shown in Fig. 6 of \citet{loll13}.  As impressive and beautiful as the full mosaic is, it is of course the detailed fine structure within the filaments, only apparent upon closer inspection, that is the hallmark of HST data.

One of the complications of observing the Crab is its dynamic nature.  Fabry-Perot and SITELLE FTS observations of the Crab show an overall expansion with a full velocity range of $\pm 2000 \kms $, although $\pm1250 \kms$ encompasses the bright filament emission (\citealt{lawrence95, martin21}, Ding et al., in preparation).  This manifests itself in a complicated and spatially variable proper motion of the filamentary structure as it expands, with observed values in excess of 0.3$\rm \arcsec ~ yr^{-1}$ for some filaments that are moving primarily tangent to the line of sight \citep{nugent98, schaller02}.  As multiwavelength data sets are obtained over time, it becomes increasingly difficult to make accurate comparisons on small spatial scales because of this motion.

This confusion now extends to JWST, where JWST cycle 1 NIR and MIR imaging observations of the Crab have been obtained \citep{temim24}, but where no contemporaneous HST-resolution images have been available for comparison.
Surprisingly, a search of the MAST archive indicates that there have been no HST ACS or WFC3 emission line images obtained of the Crab Nebula since the \citet{loll13} WFPC2 mosaic data dating to 1999 -- 2000 (although there have been continuum images showing changes in the synchrotron nebular structure in the region nearest to the pulsar). Throughout this paper, we will refer to these earlier WFPC2 data as Epoch 1. 

In this paper, we report a new epoch of HST imaging of the Crab Nebula using WFC3 and filters similar to the WPPC2 filters used by \citet{loll13}, as well as new data in \hb\ and two medium-width continuum-dominated bands.  We will refer to these data as Epoch 2.  We compare these new data both to the Epoch 1 WPFC2 data as well as the recent JWST imagery reported by \citet{temim24}.

\begin{figure*}
\center
\includegraphics[width=0.85\textwidth]{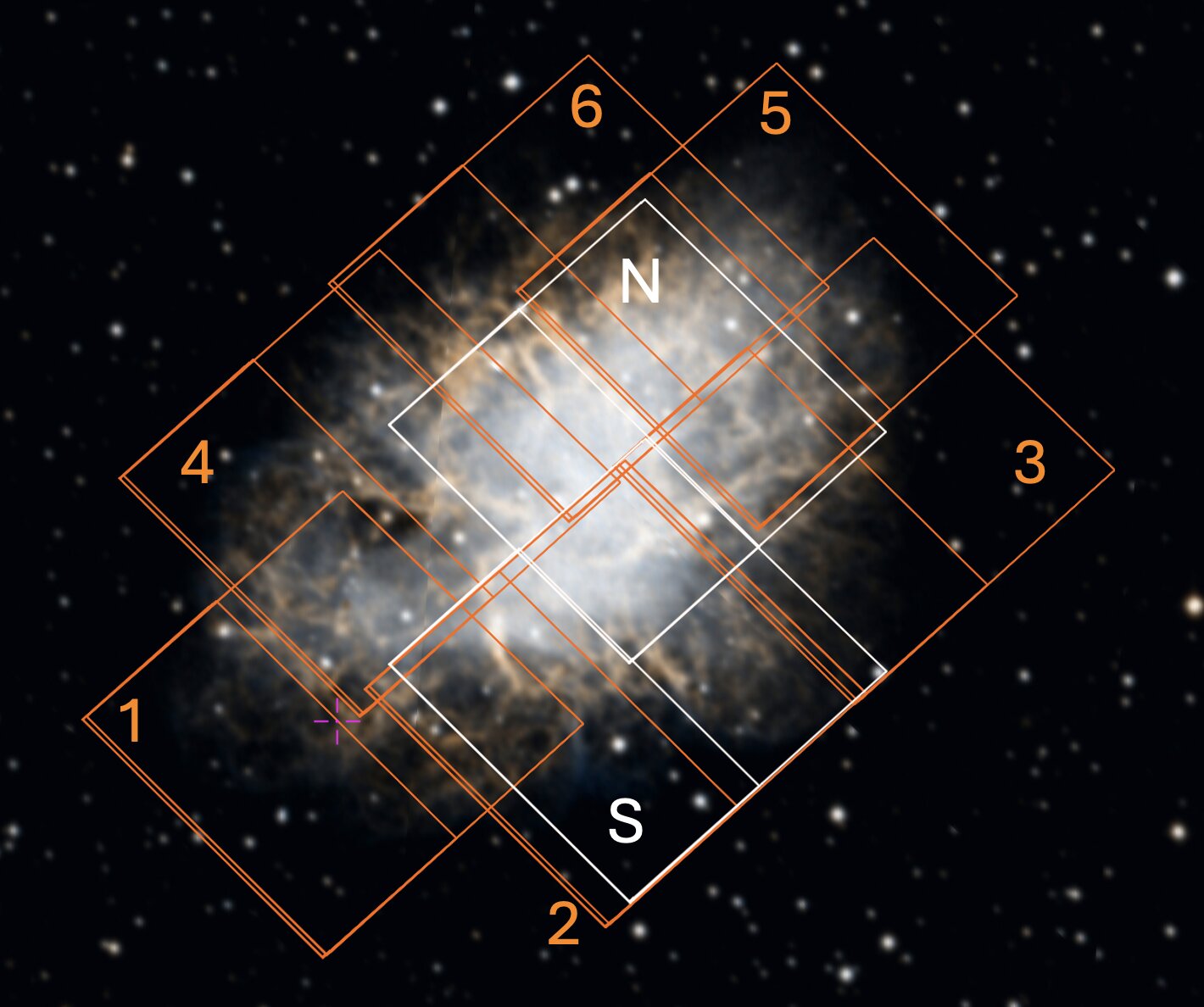}
\caption{  This figure shows the WFC3 field locations and overlaps projected onto a color DSS image, as displayed in the MAST portal.  The field numbers indicate the fields as listed in Table~\ref{fields}.  The two white boxes show the positions of the two H$\beta$ fields that cover the north/central and south portion of the nebula but do not align exactly with any of the six main mosaic component fields. In this and all figures that follow, North is up and East to the left.
\label{wfc3_fields}
}

\end{figure*}

\section{Data and Data Processing \label{sec:observations}}

The data sets used in this paper can be found using the following DOI:
\dataset[10.17909/per1-hf17]{http://dx.doi.org/10.17909/per1-hf17}.

\subsection{Epoch 2 HST WFC3 Imaging}   

The new data for this paper were all obtained under program ID 17500, a 36 orbit HST Cycle 31 program.  Six separate but overlapping WFC3 field positions (see Table~\ref{fields} and Fig.~\ref{wfc3_fields}) were observed in five filters (see Table \ref{filters}), in a sequence of exposures that required five orbits per field, for a total of 30 orbits.  The filters were selected to approximately duplicate the WFPC2 filters used in the earlier Crab HST mosaics, as presented by \citet{loll13}.  The WFC3 filters included F502N, which captured \oiiiL, F610N which isolates \oiL, and F673N which covers \SiiLL.  Because the narrow-band filters also pass significant continuum emission from the synchrotron nebula excited by the pulsar, and because the synchrotron nebula itself is of considerable interest, we also observed in two medium-band filters, F547M and F763M.  Neither of these filters is completely devoid of emission lines, but both are dominated by the synchrotron emission, and additionally can be used to subtract synchrotron from the emission line images, resulting in relatively clean emission line maps.

\begin{figure*}
\center
\includegraphics[width=0.9\textwidth]{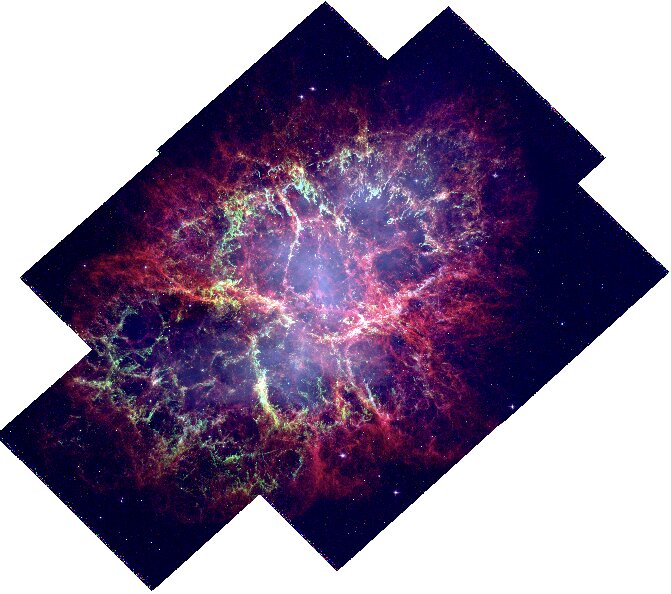}
\caption{ This figure shows a color mosaic of the six WFC3 fields using the three primary emission line filters: F502N (red), F673N (green) and F631N (blue), sampling \oiiiL, \siiL, and \oiL, respectively. Colors here were chosen in homage to the earlier mosaic of \citet[][see Fig. 6]{loll13}, and log scaling is used to compress the dynamic range.  No continuum subtraction has been done in this figure, so the synchrotron continuum being passed by the narrow filters causes the diffuse interior glow. 
\label{loll_comp}
}
\end{figure*}

Two additional fields covering the bright north/central filaments and the southern regions observed with JWST MIRI \citep{temim24} were observed with F487N (H$\beta$) to obtain information of the pure hydrogen emission for comparison to the other emission lines.  While \ha\ is of course much brighter, the proximity of the \NiiLL\ lines to \ha, the filter widths, and the kinematic spread of the filaments make the use of the relevant \ha\ filters in both WFPC2 and WFC3 filters totally useless for isolating hydrogen emission.  (Indeed, this is why \ha\ images were not obtained by \citet{loll13}, and why the \oiii, \oi, and \sii\ filters were chosen in that study.) The Crab is moderately reddened (typical E(B$-$V) values are 0.3 -- 0.5 mag), which further depresses \hb.  We observed each of the two WFC3 F487N fields with three orbits (7800 s exposure times), which was sufficient to obtain reasonable data on the bright filaments in the observed regions.  

The data for each field were obtained consecutively, thus minimizing any time-dependent changes within each field.  The bulk of the data were obtained over approximately one month from mid-February to mid-March 2024 (see Table~\ref{fields}), a time frame for which few if any observable changes would be expected in the emission filaments.  However, the inner portions of the synchrotron nebula closest to the pulsar are variable on week-to-week timescales \citep{hester02}, and so could vary somewhat between fields.  
 
Finally, the exposures for F547M (Field 1 and Field 2) and F763M (Field 1) were impacted by loss of tracking during the initial observations, requiring reobservation at a later time. The reobservation of F547M for Field 2 was successful, but the Field 1 data for F547M and F763M also failed a second time; the observing window for the Crab closed prior to attempting these filters again, and so the full region covered by Field 1 does not currently have successful continuum images for comparison. As seen in Fig. \ref{wfc3_fields}, roughly half of Field 1 overlaps with Fields 2 and 4, thus minimizing the impact from these missing data in the outer (fainter) synchrotron nebula in Field 1.  
 
Table~\ref{filters} summarizes the exposure information used for each of the six mosaic fields (five filters) as well as the two \hb\ fields.  All data were dithered by either two or three steps, dictated partially by the way the exposures laid out in the orbit planning, and postflash was used to at least partially mitigate the effects of charge transfer inefficiency in the WFC3 detectors.

\begin{figure*}
\center
\includegraphics[width=0.9\textwidth]{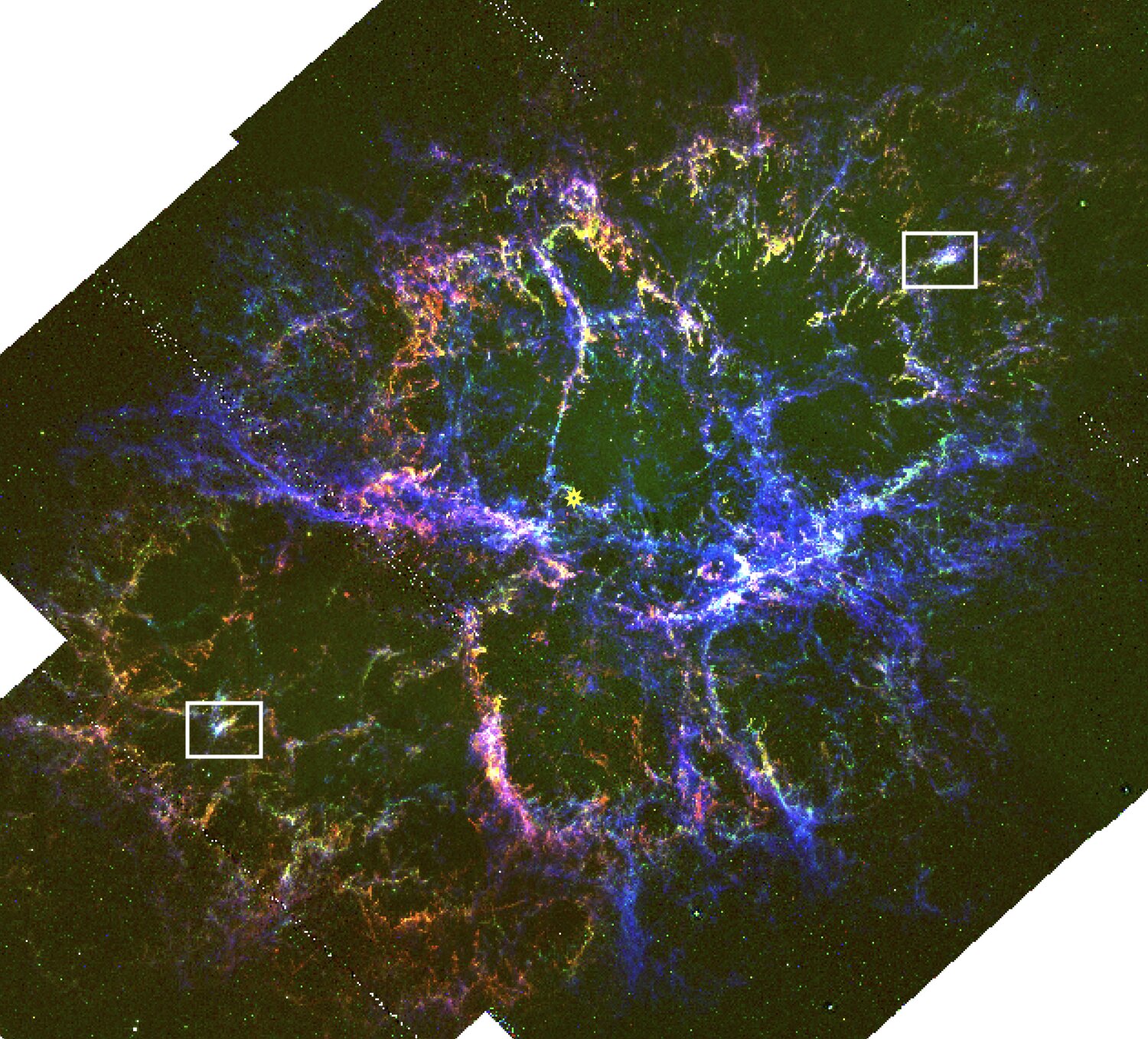}
\caption{This figure shows the primary emission line mosaic after scaling and subtracting the underlying continuum emission. In this figure and many of the three-color figures that follow, we show  \sii\ in red, \oi\ in green, and \oiii\ in blue, all shown with linear scaling. Viewing the data this way highlights the variations in relative line intensities across the filamentary structure. The two white boxes highlight two regions discussed later in the paper and the position of the pulsar is noted by a small yellow star symbol. Appendix Fig. \ref{A1fig} through Fig. \ref{A4fig} show enlargements of several sub-sections of this image to show detail. 
\label{subtracted_color}
}
\end{figure*}

\subsection{Epoch 1 Archival HST WFPC2 Imaging} 

For comparison purposes, we downloaded and reprocessed the earlier WFPC2 data (HST program 8222; Hester, PI) from MAST
for use in this project, aligning them to the same astrometric solutions (using Gaia stars) to make accurate comparisons with Epoch 2 and  other data sets. 

One complication for the comparison between WFPC2 and WFC3, however, is that even though the emission lines being captured by the filters are similar, the filter widths (and hence velocity coverage) of the WFPC2 and WFC3 filters are different. In Table~\ref{comparison}, we assess this situation: assuming an approximate $\pm 1250 \kms$ velocity spread for the bulk of the bright filaments in the Crab \citep{lawrence95}, it is clear that the WFC3 filters capture significantly more of the blue-shifted (near side) emission than the earlier WFPC2 filters. For positional comparisons, this should have little or no effect, but for any comparisons of relative brightness between the two epochs, a careful assessment of the kinematics will be required.

\vspace{0.5in}

\section{Analysis}

\subsection{Astrometry and Creation of Mosaic Images}


Mosaic images were constructed from the WFC3 data using the \texttt{AstroDrizzle} software package included in \texttt{DrizzlePac 3.0} \citep{Anand2025}. For each filter, all available exposures were combined to produce a single, deep, drizzled mosaic with a final pixel scale of 0.04 arcsec/pixel. Exposures affected by loss of spacecraft guiding or other anomalies were excluded to ensure image quality. Sky subtraction across the mosaic was performed using the \texttt{globalmin+match} algorithm, which used the mode of sky values measured from individual input images. The \texttt{skylower} and \texttt{skyupper} parameters, which define the range of pixel values used for sky estimation, were determined from each filter’s input images and applied during mosaic creation.  Astrometric alignment was refined using sources from the Gaia DR2 catalog \citep{Gaia2018,Lindegren2018}, ensuring accurate registration across all filters.

A full color mosaic of the six WFC3 fields from Epoch 2 is shown in Fig.~\ref{loll_comp}.  This Figure adopts the color assignments used by \citet{loll13}, where \oiii\ is shown in red, \sii\ in green, and \oi\ in blue, but in the remaining figures below where all three emission lines are combined, we adopt a different color assignment strategy. Despite the dithering, some residual effects from cosmic rays are apparent in the chip gap regions of each field in the pipeline-processed data files available from MAST\footnote{\url{https://archive.stsci.edu}}.

\subsection{Removal of Continuum from Epoch 2 Emission Line Images \label{sec:image_subs}}

The narrow filters isolate emission lines but are still broad enough to pass some of the synchrotron continuum emission as well (cf. Table~\ref{comparison}). For the assessments below using the new WFC3 data, we want to have reasonably pure emission line images. 
The F547M image is nearly line free and makes the best filter for scaling and subtracting from the narrow filters.  Scaling by the relative widths of the filters before subtracting gets close, and then small tweaks allow for the best scalings for final subtraction. Fig.~\ref{subtracted_color} shows a wide field color version of the continuum-subtracted WFC3 emission line mosaic data for the three primary emission filters.  This figure highlights the global spatial relative line variations. The white boxes highlight two regions discussed in Sec.~\ref{sec:surprising}. Since it is the fine-scale and resolution of the filaments that is the strength of HST imaging with WFC3, we show several subregions from this mosaic in more detail in Appendix Figures \ref{A1fig} through \ref{A4fig}.

F763M also primarily captures continuum but many of the bright filaments are also apparent in these data.  Judging from published spectra \citep{dennefeld83, macalpine89}, the primary emission line within the F763M filter bandpass is \nicL, although \oii\ $\lambda\lambda$ 7320,7330 can also contribute at a lower level. Any quantitative comparisons of F547M and F763M images to study the synchrotron spectral index will need to avoid filament regions.

\section{Results}

\subsection{General Motivations}

Optical emission from the Crab is dominated by two sources: 1) the diffuse continuum-emitting synchrotron nebula, which is energized by the pulsar, and 2) the expanding network of ejecta filaments that show a range of emission-line properties.  The ejecta filaments are themselves excited and ionized by the synchrotron emission, which varies in strength and hardness ratio as a function of position outward from the pulsar.  Furthermore, photoionization modeling of optical and near infrared observations indicate the chemical composition can vary substantially from filament to filament \citep{henry82,pequignot83, henry86, macalpine08}.

The filaments often show evidence of a range of densities, with dense cores and lower density surrounding material, as evidenced by both dust shadows seen against the background of synchrotron emission (for filaments on the near side) and variable ionization states on small spatial scales. In other young historical SNRs like Cas A and Kepler's SNR, knots and filaments are known to vary on timescales of a decade, but variability the fine scale structure or brightness of Crab filaments has not been studied to date.

These new HST WFC3 images open up a plethora of potential comparisons with other data sets.  With the large time separation from the earlier HST WFPC2 imagery and the similarity of the filters used, time variable changes on a 20+ year time baseline can be studied. These include not only expansion of the filamentary structure but also potentially any changes in relative line intensities, structures within the filaments and/or synchrotron nebula, and possible changes in the dust shadows cast by filament cores as viewed against the sychrotron background.\footnote{The caveat of the different filter widths must be noted, however.} As in the Epoch 1 study, the emission line filters were chosen because they are able to isolate the emission from individual ions (\oiiiL, \oiL, and \siiL) even given the expected expansion velocities. Comparisons between these three ions provides qualitative information about filamentary ionization structures (e.g., \oiii\ to \oi\ and \oiii\ to \sii, with some uncertainty from possible variable abundances with position in the latter case).

Additionally, there are comparisons within the WFC3 data themselves that are of interest. Recall that the spatial resolution of WFC3 is roughly a factor of two higher than even the WFPC2 data, potentially resolving additional structure within the ejecta filaments. We also obtained F487N data for the bright central filaments, permitting comparisons against a pure hydrogen line (\hb). Finally, two separate medium-band filters that are dominated by synchrotron emission were obtained in our recent data, providing a look into any changes in synchrotron spectral index within the HST range.

As stated earlier, these new data are nearly contemporaneous with recent JWST imagery as well \citep{temim24}, making comparisons between these two powerhouse instruments more straightforward. However, one still needs to contend with varying spatial resolution with wavelength and the need to separate emission line and continuum components of the emission.

Below we make an initial foray into these comparisons while acknowledging that more quantitative assessments will be needed for detailed comparisons with computer modeling of the various emission components.

\subsection{HST Two Epoch Comparisons}

\subsubsection{Proper Motion Between Epochs \label{sec:proper_motion}}

The Crab Nebula is known to be expanding away from the point of the explosion in AD 1054.  Furthermore, it is known that the filamentary expansion is actually accelerating due to the outward pressure from the pulsar wind nebula \citep[cf.][and references therein]{nugent98}. This causes the positions of the filaments to change over time at a level that is observable even from the ground \citep{trimble77, wyckoff77, nugent98, martin21}.  The proper motion of any particular filament will depend not only on the space velocity of that filament, but the projection of this velocity onto the plane of the sky. Hence, the proper motion is expected to change systematically across the face of the remnant. For filaments moving in the plane of the sky, this motion is roughly 0.3\arcsec\ $\rm yr^{-1}$ \citep{nugent98}. One of our primary motivations for reobserving the Crab with HST was to obtain a data set more contemporaneous with current epoch multiwavelength data sets and to enable accurate and detailed multiwavelength comparisons even at the finest spatial scales and for individual knots and filaments, which of course is the unique strength of HST.

\begin{figure*}
\center
\includegraphics[width=0.9\textwidth]{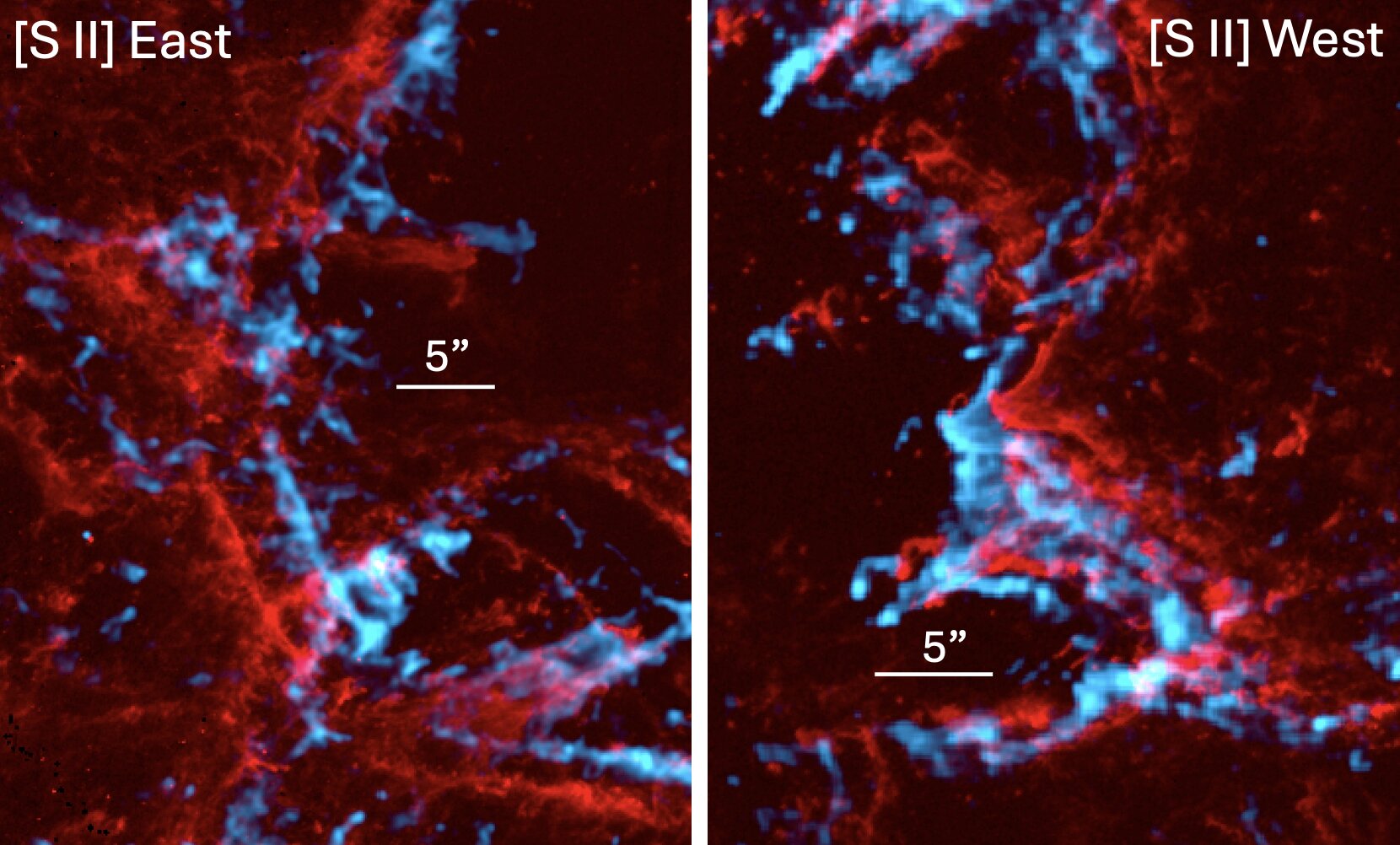}
\caption{This figure shows enlargements of the two epoch data for \sii. Cyan shows the 2000 WFPC2 data and red shows the 2024 WFC3 data. (left) A region toward the east side of the Crab and (right) for a region toward the west side.  These regions were chosen because the motion is primarily in the plane of the sky at these positions. The outward motion is obvious. Both epochs show similar structure, with few if any discernable changes in relative intensity.
\label{propmotion}
}
\end{figure*}

Detailed ground-based reassessment of Crab expansion characteristics covering the entire nebular structure has been performed recently using the imaging Fourier transform spectrometer SITELLE and MegaPrime/MegaCam imager on the Canada–France–Hawaii Telescope, thus obtaining detailed kinematic information \citep{martin21,martin25}.  Rather than duplicate this extensive analysis, in Fig.~\ref{propmotion} we simply show two examples using the \sii\ data from the two epochs.  The HST spatial resolution reveals incredible detail that could be applied anywhere in the coverage area to determine accurate proper motions of any particular filament, knot, or region of interest within the Crab's structure, as might be desired in future analyses. 

In addition to the obvious proper motion, we have inspected a number of regions looking for obvious changes in the relative intensities of neighboring knots and filaments in the the \oi\ and \sii\ images. (The \oiii\ emission is more diffuse at HST resolution and harder to compare.) From extensive  visual inspection, no significant changes in the morphology or intensity of bright knots has been found that cannot be attributed to differences in the filter widths between the two epochs of data (cf. Table~\ref{comparison}). This stands in contrast to other young galactic remnants such as Kepler and Cas A where individual knots and filaments do show variability on roughly a decade timescale. This difference is likely due to the differing ionization mechanisms: knots in Kepler or Cas A are impulsively heated by shock encounters while the Crab's filaments are bathed in the synchrotron emission, thus providing an ongoing (and roughly constant) source of energy input.

\begin{figure*}
\center
\includegraphics[width=0.9\textwidth]{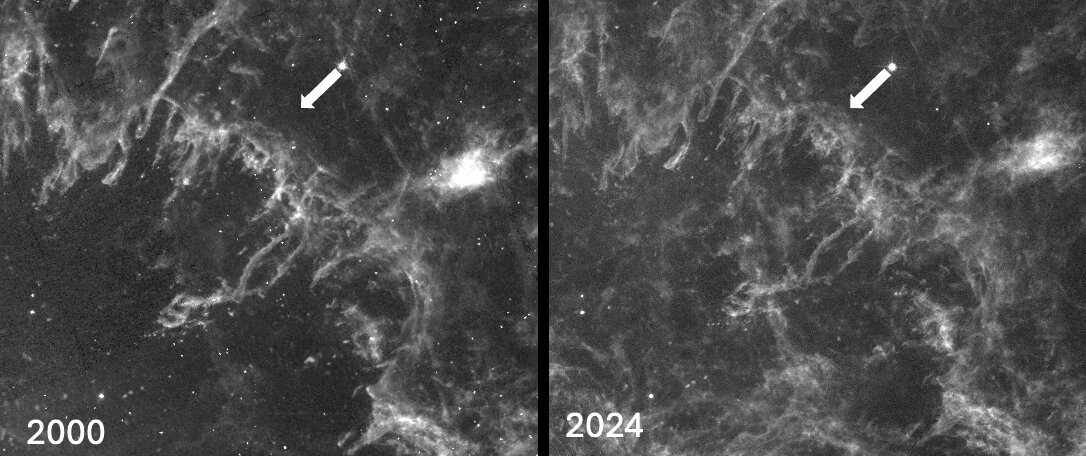}
\caption{This figure shows the two epoch comparison of \oiii\ emission for Epoch 1 (left) and Epoch 2 (right).  A number of the Rayleigh-Taylor `finger' filaments are within this region.  The fiducial white arrow is the same length and at the same position in both panels, and the general outward motion of the filaments in this region (toward the upper right) can be judged relative to this arrow.  The 2024 (WFC3) image shows a more extensive network of faint \oiii\ filaments especially visible over the lower left quadrant of the right panel.  This is due to the broader WFC3 F502N filter that captures more of the blue-shifted (near side) \oiii\ emission.
\label{RT-comp}}
\end{figure*}

In addition to brightness, it is also clear that the morphology of the filaments is essentially identical.  In particular, the R-T fingers have just moved outward with the other filaments, without stretching or changing shape (see Fig.~\ref{RT-comp}).  As a general observation, the propensity of the WFC3 F502N filter to capture more of the front side (blue-shifted) emission compared to the earlier data is particularly obvious as a significantly more extensive network of faint filaments seen in projection toward the central regions, also visible in Fig.~\ref{RT-comp}.

\begin{figure*}
\center
\includegraphics[width=0.9\textwidth]{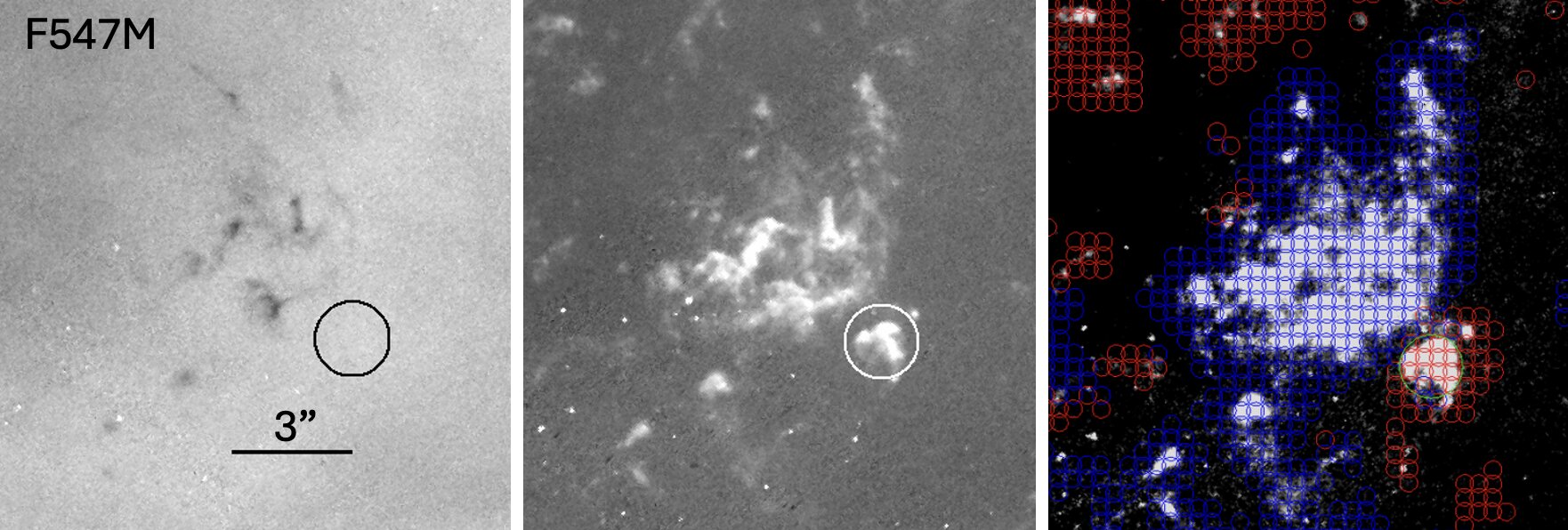}
\caption{This figure shows a small region southeast of the pulsar that contains a number of dust shadows visible in the F547M continuum band (left); for comparison, the same region is shown from the \sii\ frame (middle). These faint filaments are barely visible in \oi, which is not shown.  Clearly the dust shadows are associated with some of these filaments, but not necessarily the brightest or most well-defined filaments, such as those highlighted by the circle. Indeed, the panel at right shows color-coded velocity information derived from SITELLE that confirms both that the main body of filaments that show dust absorption are blue-shifted, and that the bright filament in the circle is redshifted and hence behind most of the synchrotron emission.
\label{dust1}
}
\end{figure*}

\subsubsection{Dust Shadows \label{sec:dust}}

As seen in earlier data (including ground-based data), dark regions are visible from near-side knots and filaments that are seen in projection against the synchrotron emission.  These dark regions arise due to dust absorption within certain ejecta filaments \citep{fesen90, grenman17}. Of course, more global estimates of both the distribution and mass of dust in the Crab have been made using infrared observations \citep{gomez12a, owen15, delooze19, temim24}, finding that dust seems to be distributed largely within the regions of the bright inner `cage' filaments surrounding the pulsar.  The HST data permit us to not only see the distribution of the densest regions of dust on a filament-by-filament basis, but to resolve significant structure in the dust, from knots and filament cores to a more diffuse component at some locations that has not been fully appreciated in earlier data.

Which filaments show dust shadows depends on a number of factors, including not only whether they contain dust or not, but how this dust is distributed and whether a given filament is on the near-side of the Crab so that its dust can be seen against the background synchrotron nebula. Fig.~\ref{dust1} shows a collection of relatively faint filaments southeast of the pulsar that show relatively dark dust shadows.  While the dust is clearly associated with these filaments, it does not appear that the brightest optical filaments necessarily produce the darkest dust shadows. Hence, central filament dust density is likely not strongly coupled to filament brightness in the optical lines.  
A faint region of more diffuse emission just above center in the figure has a clear but more diffuse shadow counterpart as well.  Thus, it is also clear that dust is not confined just to the dense cores of clumpy ejecta, but can be more generally distributed.

\begin{figure*}
\center
\includegraphics[width=0.9\textwidth]{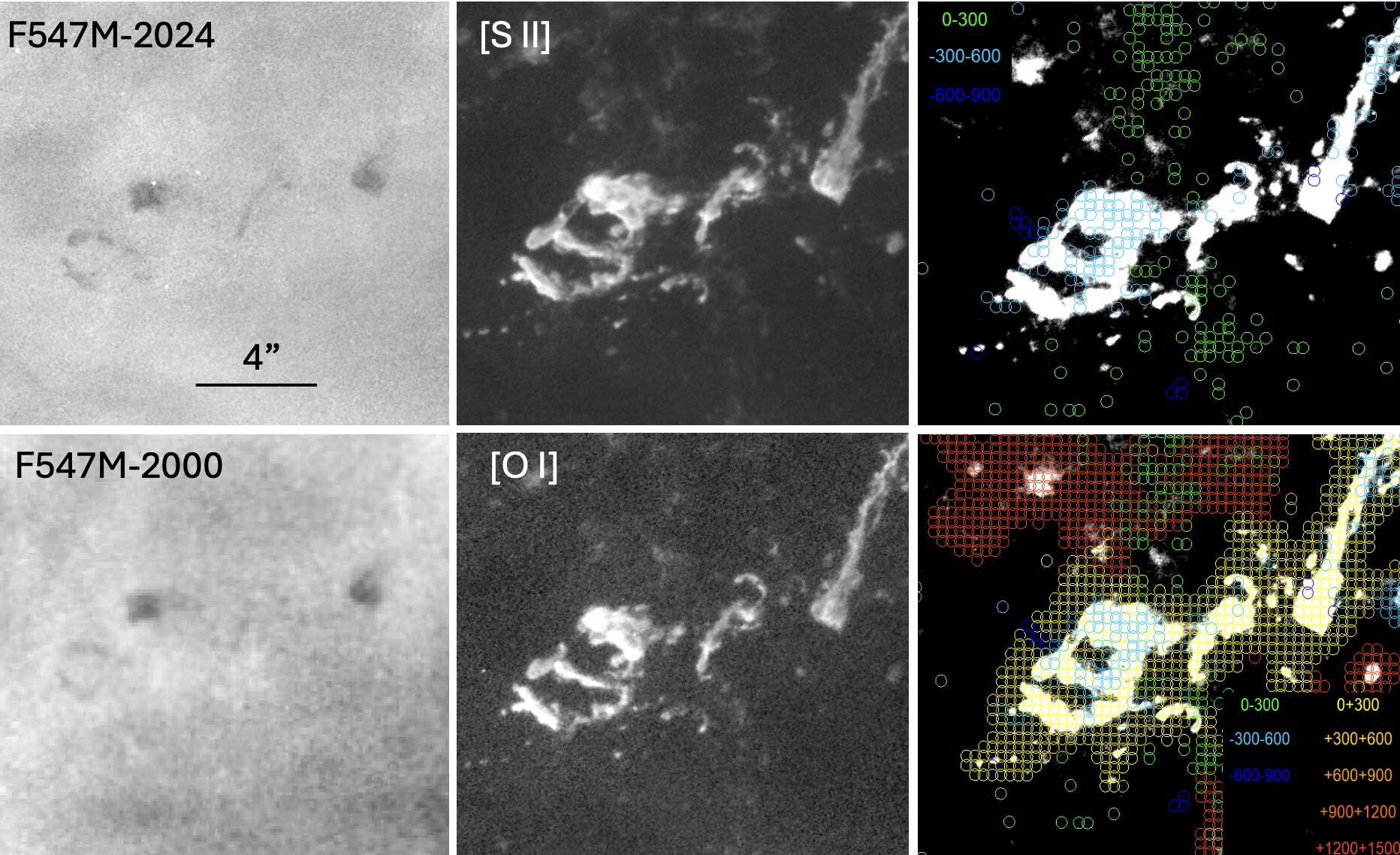}
\caption{This figure shows the dust shadows and filaments in a small region northwest of the pulsar that contains a number of dust shadows visible in the WFC3 F547M continuum band (upper left); at (lower left) the 2000 WFPC2 continuum image is shown for comparison. Recalling that the resolution of WFC3 is better than WFPC2, there is no hard visual evidence for changes in the dust absorption. The upper middle panel shows the 2024 \sii\ image; the lower middle panel shows the 2024 \oi\ image. The panels at right show color-coded velocities from the SITELLE data, where the top panel shows only the blue-shifted emission and the lower panel shows the full velocity range.  This region is notable for containing several Rayleigh-Taylor filaments that are bright in both \sii\ and \oi, which look nearly identical.  The heads of several of these R-T features show prominent dust absorption.   The filaments showing the darkest dust shadows appear to be only modestly blue-shifted. 
\label{dust2}
}
\end{figure*}

\begin{figure*}[b]
\center
\includegraphics[width=0.8\textwidth]{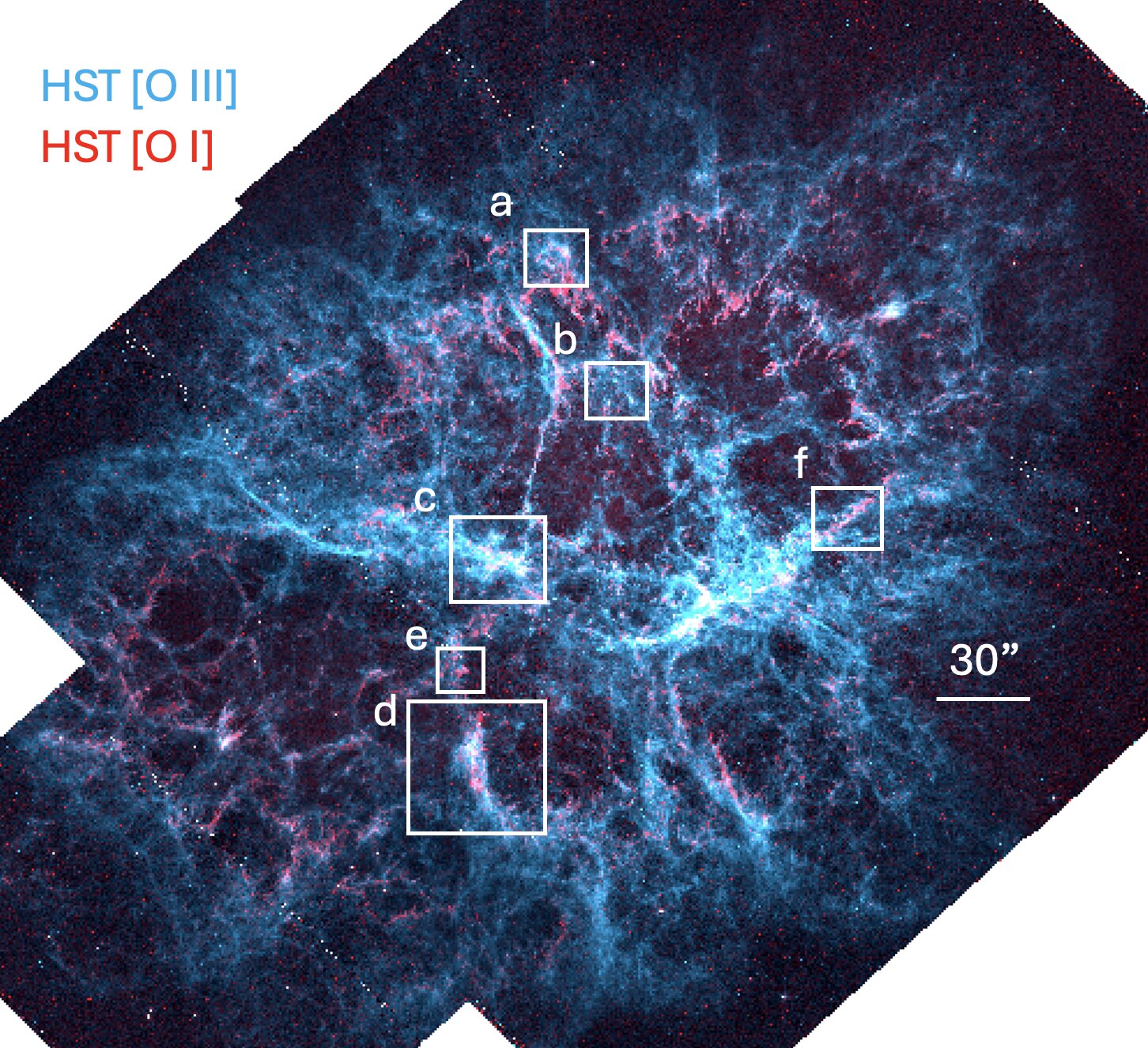}
\caption{This figure shows a comparison of \oiii\ in cyan and \oi\ in red, thus highlighting the ionization structure of the filaments.  Both images are shown on a log scale to compress the dynamic range. \oi\ arises in the compact dense cores of filaments, primarily in the inner `cage' filaments. In contrast, the morphology of the \oiii\ filaments is more diffuse and more generally distributed on the large scale.  The white boxes highlight regions enlarged in the following figure that show the finer scale structure accessible by HST/WFC3. 
\label{ionization1}
}
\end{figure*}

\begin{figure*}[b]
\center
\includegraphics[width=0.95\textwidth]{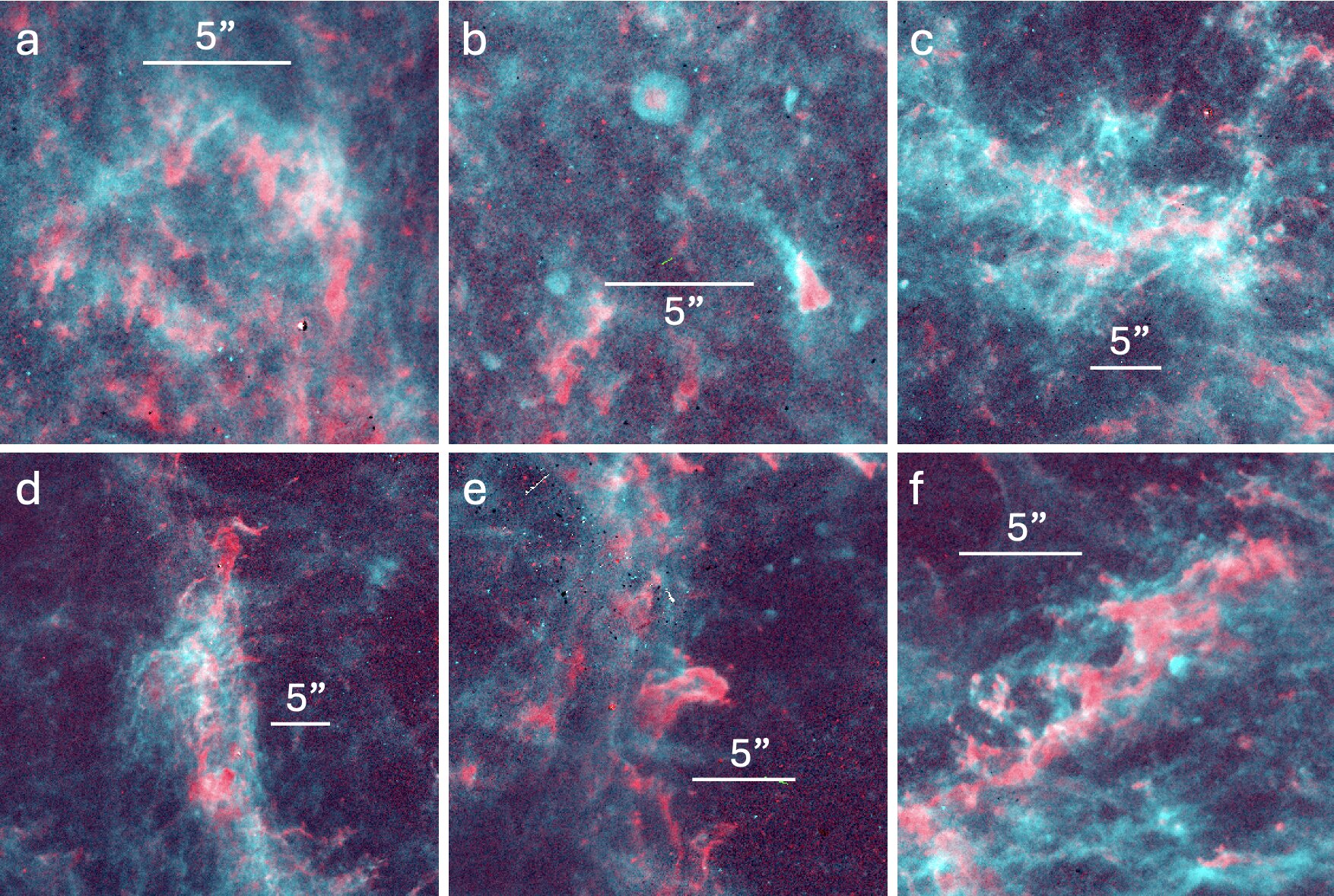}
\caption{This figure shows enlargements of the six regions highlighted in the previous figure, to show the detail.  The colors are the same but scaling has been adjusted slightly to avoid saturation and show detailed structure.  
\label{ionization2}
}
\end{figure*}

In the right panel of Fig.~\ref{dust1}, we use SITELLE data (Ding et al., in preparation) to inspect the velocity structure within this region.  It is clear that the filaments causing shadows are indeed blue-shifted in the range -600 to -750 $\kms$, and hence on the near side of the nebula.  The brightest filament in the middle panel of the figure does not show any shadow, but SITELLE shows it to be red-shifted by $\sim$500$\kms$.  This demonstrates the complex relationship between appearance in the imagery versus the complex velocity structure along any given line-of-sight through the nebular structure. 

Fig.~\ref{dust2} shows a small region northwest of the pulsar that contains several Rayleigh-Taylor filaments with prominent dust shadows, especially from the dense `heads' of the R-T finger filaments \citep{blair97, sankrit98}, as seen in the left panels.  In this case, the low ionization \oi\ and \sii\ emission are both bright (middle panels), and the morphology of the filaments is nearly identical in both lines.
At the bottom left in this Figure we show the Epoch 1 WFPC2 F547M data for comparison.  Within the constraints of the differing resolution and changing background, there is little evidence of changes in the dust absorption from these filaments. 

The right two panels of Fig.~\ref{dust2} show the velocity structure of this region as seen in the SITELLE data.  The overall velocity structure in the region is complex, but the bright R-T finger emission is dominated by only very modestly blue-shifted (0 to -300 $\kms$) emission.  Given the outward proper motion of this region (cf. Fig.~\ref{RT-comp}), it seems likely that these filaments are not fully on the near side of the nebula but rather viewed partially within the synchrotron emission itself.  Since the synchrotron nebula is not fully behind these filaments, their dark shadows are thus somewhat surprising, likely indicating quite high densities of dust in these filaments.

\subsection{WFC3 Ionization Structure Assessments \label{sec:ionization_structure}}

In this section, we highlight some of the ionization structures visible within the new WFC3 data.  The structure of the Crab's ejecta filaments is complex, with dense cores and lower density surrounding material. This in turn makes their ionization structure complex; the filaments are ionized by a diffuse bath of surrounding synchrotron radiation, which then interacts with this density structure. Furthermore, the recombination rate is proportional to density, adding to the difficulties in modeling the filament emissions.  Finally, the intensity of the synchrotron radiation varies with position as well.  Thus, interpretting comparisons between filaments can be difficult. HST spatial resolution allows us to view the structure of individual filaments in detail.  By carefully selecting filaments that are not too confused by their surroundings, one can hope to separate out some of this complexity.

\begin{figure}[b]
\center
\includegraphics[width=0.75\textwidth]{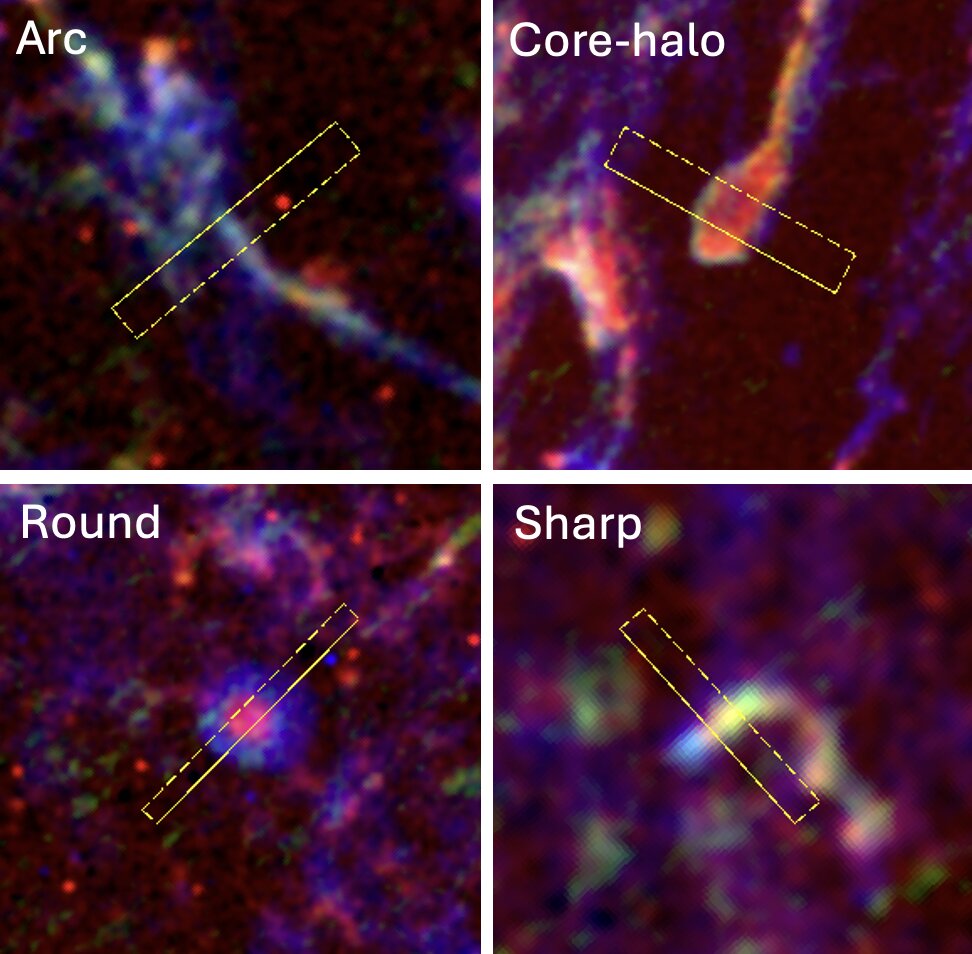}
\caption{This figure shows four regions selected for cross-cuts to study fine-scale ionization structure.  The background image has \oi\ in red, \hb\ in green, and \oiii\ in blue.
\label{ion-reg}
}
\end{figure}

The ionization structure of the ejecta filaments was revealed by earlier HST/WFPC2 narrow-band imagery of the Crab Nebula \citep{hester96, blair97, sankrit98, loll13}. The overall structure of many of the filaments was shown to be due to Rayleigh-Taylor (R-T) instabilities occurring at the interface between the synchrotron nebula and the surrounding ejecta, followed by Kelvin-Helmholz (K-H) instabilities shearing the sides of the ``fingers'' \citep{hester96}. The R-T filaments generally pointed inwards, and their morphologies ranged from sharply defined linear features to shredded wisps of emission.  Between these extremes were several filaments showing a ``core-halo'' structure where the R-T filament heads retained their bulbous shape, and were connected back to the synchrotron nebula-ejecta interface, sometimes referred to as the `cage.'

The ionization structures of individual filaments, traced by \oiii, \sii, and \oi\ were found to be strongly correlated to the morphologies, and could be reproduced by photoionization models assuming cylindrical symmetry and the Crab's synchrotron spectrum as the ionizing source \citep{sankrit98}. On a larger scale, and particularly in the central ``high-helium'' band, the lower ionization emission lines (\oi\ and \sii) show a more structured, clumpy appearance compared with the higher ionization \oiii\ emission line image \citep{blair97}. This is largely a density effect.

Fig.~\ref{ionization1} shows this ionization structure on a global scale, with \oiii\ in cyan and \oi\ in red (and thus white representing where both ions are bright). Since both ions are from O, there is no confusion from abundance variations. Ionizing from $\rm O^{+}$ to $\rm O^{++}$ requires 35.12 eV while neutral O becomes  ionized to $\rm O^{+}$ at only 13.62 eV. Thus, 
\oi\ knots and filaments are indicative of dense, neutral filament cores that are shielded from the synchrotron ionizing emission.  In contrast, \oiii\ appears `fluffy' and is much more generally distributed even well out into the outer nebular filaments. We note that significantly more near-side (blueshifted) \oiii\ is passed by the WFC3 F502N filter compared with the earlier WFPC2 image, causing some modest confusion by foreground emission.

In Fig.~\ref{ionization2}, we show enlarged regions of the six areas indicated by white boxes in Fig.~\ref{ionization1}. This figure shows the detailed structure of selected filaments.  In some regions, \oi\ cores appear within extended diffuse \oiii, revealing the ionization structure within individual filaments.  However, there are many examples where the two ionization states are spatially separated from each other, which again is largely a density effect with higher ionization corresponding to lower density (more diffuse) gas.  For future modeling, selecting individual knots and filaments that show clear ionization structure will be the most useful.

\begin{figure}[b]
\center
\includegraphics[width=0.95\textwidth]{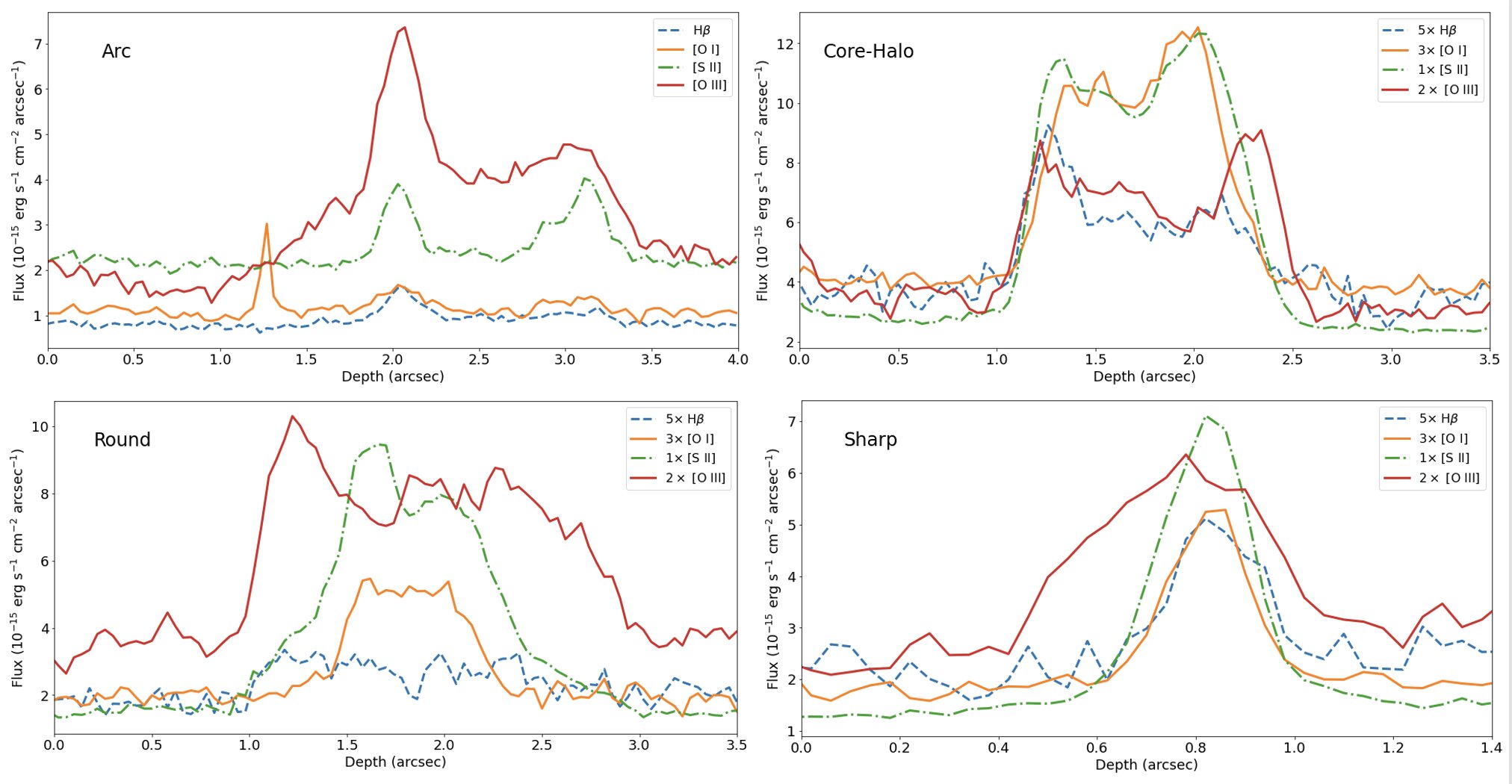}
\caption{This figure shows cross-cuts of selected features in various emission lines, showing the resolved ionization structure.  The more extended nature of the \oiii\ emission is readily apparent. 
\label{ionization3}
}
\end{figure}

We now briefly assess the ionization structure of individual filaments in the Crab Nebula using the WFC3 images, which provide three advances over the earlier studies: (i) the angular resolution is a factor of two higher, (ii) the F502N filter bandpass is significantly broader and transmits the \oiii\ emission from almost the entire range of blue- and red-shifted velocities expected, and (iii) the availability of a sufficiently deep H$\beta$ image over a significant fraction of the nebula that allows a direct comparison of the oxygen and sulfur lines to a hydrogen line.

We have made cross-cuts for a selected set of filaments and emission features to demonstrate the variations in ionization structures and relative line strengths present in the Crab. All of these regions have been selected to lie within the FOV of the H$\beta$ image while also being reasonably isolated from line-of-sight contamination by unrelated filaments. The yellow rectangles in Fig.\ref{ion-reg} show the positions of the cross-cuts, and Fig.~\ref{ionization3} shows plots of the line fluxes against distance along these cross-cuts. The fluxes have been obtained by summing the count-rates across the width of the cross-cuts and converting them to physical units using the equation given in \S9.4.3 of the WFC3 Instrument Handbook (Version 17.0, December 2024). For this conversion, we assume that emission at a single wavelength contributes to the flux, and the following throughput values: F487N -- 0.25, F502N -- 0.24, F631N -- 0.23, and F673N -- 0.24. They have not been corrected for interstellar reddening. In the plots, the flux units are erg~s$^{-1}$~cm$^{-2}$~\arcsec$^{-1}$. The conversion from image to angular scale is 1 pixel = $0\farcs04$. At the distance of the Crab this corresponds to 1.2$\times\,10^{15}$~cm.

The ``Core-Halo'' filament is the one designated filament ``F'' in \citet{sankrit98}, and the ``Sharp'' filament is similar to filament ``D'', which itself is outside the FOV of the WFC3 H$\beta$ image. The difference in ionization structure between these two types of filaments was ascribed to differences in the magnetic field strengths, whereby stronger fields prevented secondary KH instabilities from developing in the case of sharp filaments \citep{hester96}. Photoionization models of the filaments that matched the \oiii, \sii, and \oi\ emission profiles indicated peak densities of n$_{H}\approx 1100$~cm$^{-3}$ \citep{sankrit98}. In the case of sharp filaments these are expected to be the true central densities, while in the case of core-halo structures it is likely that densities are much higher in the core. 

The details of the emission line cross-cuts for the Core-Halo filament suggests that this is the case - H$\beta$ tracks the \oiii\ and both \sii\ and \oi\ show a dip in the center. The cores of these filaments may be traced by dust, and possibly by molecules \citep{graham90}.
Considering the left edge of the Core-Halo filament (towards the upper-right of the projection in Fig.\ref{ion-reg}) the locations of the peak emission are resolved -- \oiii\ peaks first and H$\beta$ about 0\farcs04\ further in. Sulfur has a lower ionization potential than hydrogen, and as expected peaks about 0\farcs07\ closer to the center than H$\beta$. The exact location of the \oi\ peak is not well defined, but it is more than 0\farcs1\ away from the \oiii\ peak. The scale of stratification in the ionization is a direct consequence of the cross-sectional density profile of the filament.

In the case of the sharp filament, the emission lines all peak at the center, but there is a noticeable stratification along the filament as seen in Fig.\ref{ion-reg}. The tip of the filament is pointing inwards and is dominated by \oiii\ emission. This suggests a much shallower density gradient along the filament than across it.

The ``Round'' filament (named simply for its circular appearance) has an ionization structure that is clearly stratified.  This filament likely represents the head of a R-T finger where the finger is pointing directly toward us.  To first order it appears to be similar to the Core-Halo filament. However, the H$\beta$ and \oi\ fluxes are significantly lower than \oiii\ and \sii\ compared with the Core-Halo filament. It is difficult to explain this purely in terms of a density gradient, and may be due to the specific viewing angle and curvature of the R-T finger that results in a longer path length through certain ionization zones.

The ``Arc'' filament, unlike the others, is not an R-T finger but rather part of an interface between the synchrotron nebula and ejecta. It resembles the regions that lie between the points of attachment of R-T fingers, being curved with the convex side towards the remnant center. The difference in the ionization structure and relative line strengths are clearly seen in Fig.~\ref{ionization3}. 
In this plot, the center of the remnant is towards the left, the brighter arc (Fig.\ref{ion-reg}) lies between about 1.0 and 2.8\arcsec\ along the x-axis and the secondary fainter feature lies beyond that. Also, for the Arc,
the fluxes are not scaled as in the other three cases. The strongest line is \oiii\ and all the lines have similar profiles, indicating that the density is relatively low, the structure is density-bounded in the transverse direction, and the line-of-sight depth through the emitting region is relatively large. 

\begin{figure}[t]
\center
\includegraphics[width=1.0\textwidth]{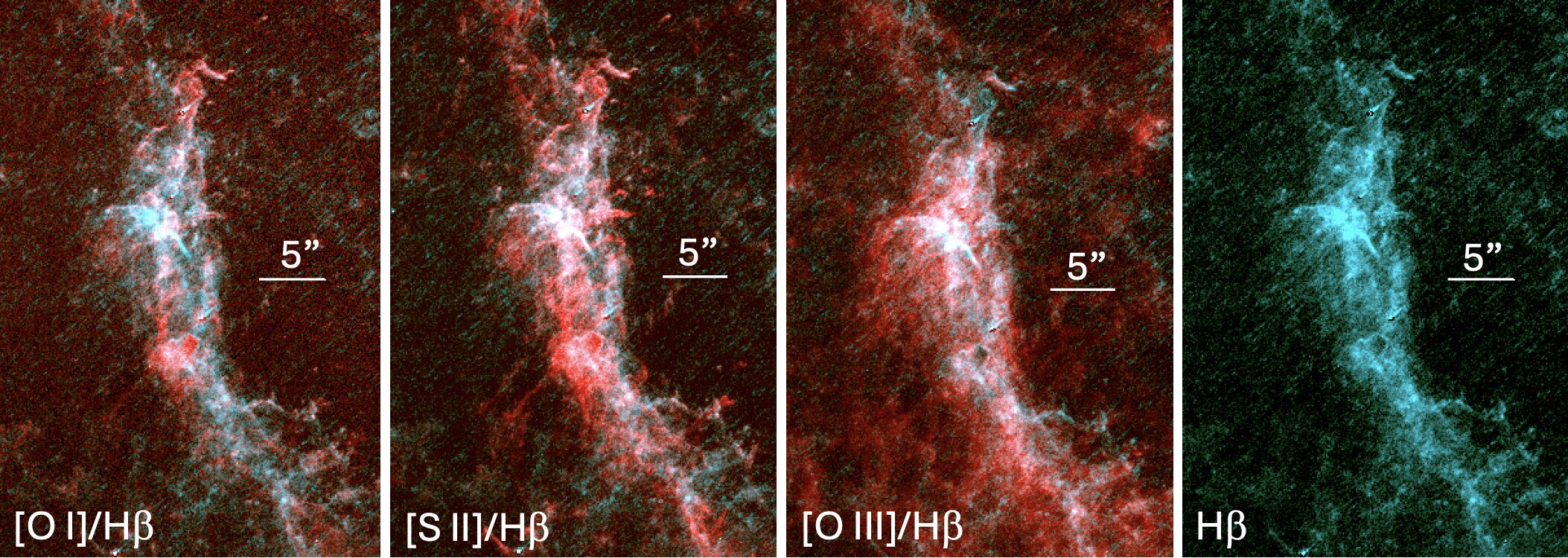}
\caption{This figure shows several panels comparing various ion lines (in red) to \hb\ (in cyan) for the bright southern filament. The \hb\ image stays the same for all panels, with the righthand panel showing \hb\ alone for context. Regions where both emissions are bright will display as white. 
\label{comp-south}
}
\end{figure}

\subsection{Comparisons to Hydrogen Image   \label{sec:composition}}

Spatially varying line intensities from spectroscopy of the Crab have been measured by a number of investigators over the years \citep[][to name a few]{fesen82,macalpine89,macalpine08}, and while it requires careful modeling to extract abundances, chemical composition is known to change dramatically from filament to filament.
Varying line intensities in imagery from ions of different elements are hard to interpret strictly as abundance variations because the local ionization, density variations, and the chemical composition could all be changing with spatial position. For the bright central filaments, we have images in \hb, and so comparisons of the various ions to hydrogen are possible.  One can expect to first order that these comparisons might highlight actual compositional changes with position, at least for the lower ionization lines like \oi\ and \sii\ that are not too different from \hb\ in ionization potential. 

Figure \ref{comp-south} shows a four-panel figure with comparisons between \oi, \sii, and \oiii\ and \hb, with the fourth panel showing the \hb\ image by itself for context.  This figure is centered on the bright southern filament, but similar figures showing two other collections of bright filaments are shown in Appendix Figures \ref{comp-north} and \ref{comp-center}.  Once again, the more extended and diffuse nature of \oiii\ compared to \hb\ is dominated by ionization differences.

This should be much less the case for \oi\ and \sii. These species both arise in dense, low ionization regions ($\rm S^0$ to $\rm S^+$ occurs at 10.36 eV while $\rm O^0$ ionizes at 13.62 eV). They tend to appear in the same filaments, but their relative line intensities change dramatically with position. Examples for two subregions are shown in Appendix Fig. \ref{abund1}.  Such comparisons are suggestive of compositional changes, but detailed assessments of composition will require careful modeling and consideration of a much broader set of elements and line strengths.

\subsection{HST/WFC3 Comparison to JWST Imaging \label{sec:JWST_comp}}

The following two subsections show comparisons between the new WFC3 data and JWST imagery from \citet{temim24}, first on a global scale and then a detailed comparison for two small regions observed with the MIRI MRS instrument. These comparisons were hampered previously by the large time separation between the earlier HST data (circa 2000) and the JWST data, which were obtained in late 2022 and early 2023.  In contrast, the new WFC3 data were obtained in early 2024, only roughly one year after the JWST data.  For our purposes, we can consider these data sets to be effectively contemporaneous, allowing direct comparison.

\subsubsection{Global Comparisons \label{sec:HST-JW-global}}

\begin{figure}[b]
\center
\includegraphics[width=1.0\textwidth]{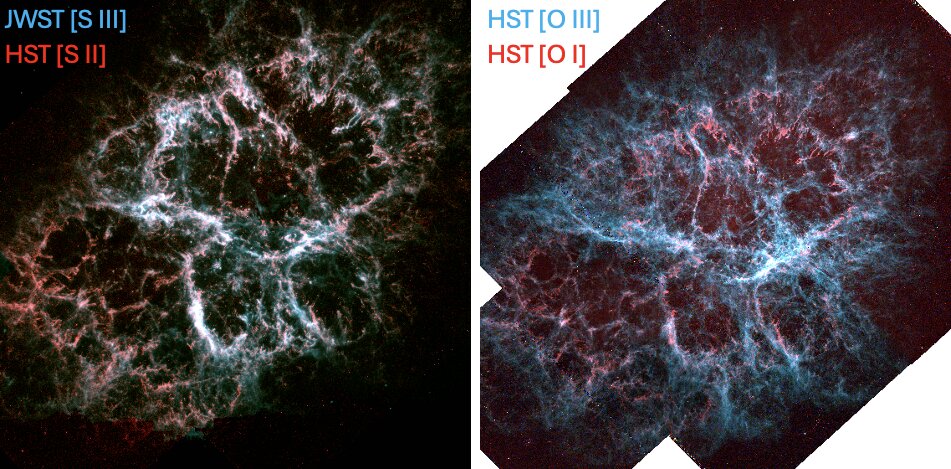}
\caption{This figure shows ionization map comparisons for JWST \siii\ and WFC3 \sii\ (left panel) and WFC3 \oiii\ and \oi\ (right panel).  In both cases, the higher ionization line is shown in teal and the lower ionization line in red. (The right panel data were shown in Fig. \ref{ionization1} but is reproduced here with adjusted display levels for ease of comparison.) 
\label{comp-s3s2}
}
\end{figure}

JWST imaged the Crab with both MIRI and NIRCam.  Unfortunately, the filters on MIRI are broad enough that images produced are not isolated on individual emission lines (and in some cases are dominated by the synchrotron emission). Using multiple filters, \citet{temim24} were able to construct an image that is dominated by \siii\ 18.7 $\mu$m, which is shown in the left panel of Fig. \ref{comp-s3s2} along with the WFC3 \sii\ mosaic.  The right panel reproduces the oxygen ionization map from Fig. \ref{ionization1} for comparison. Since these comparisons each just involve lines of one element (sulfur or oxygen), they represent clean ionization structure diagnostics.  

Although some of the same dense core/halo type structures are seen in this image, the overall appearance of the sulfur and oxygen ionization maps are quite different.  In particular, the \siii\ emission does not show the same ``fluffy'' appearance seen in \oiii, and the filaments brightest in \siii\ and \oiii\ do not match very well (see also Fig. \ref{comp-s3o3} in the Appendix). The difference is likely due to the different ionization potentials of the ions involved.  It requires 35.12 eV to make \oiii\, while it only requires 23.34 eV to create \siii.  So \oiii\ samples a lower density and more diffuse component than \siii, which is more related to the structure of denser filament complexes. \oi\ and \sii\ highlight the densest filament cores and knots.

For completeness, we show the \siii\ compared to \oiii\ images in the Appendix Fig. \ref{comp-s3o3}. Since \siii\ and \oiii\ are from different elements, it must show a combination of ionization and abundance effects, making the stark differences between these two ions hard to interpret from images alone.

\begin{figure*}
\center
\includegraphics[width=1.0\textwidth]{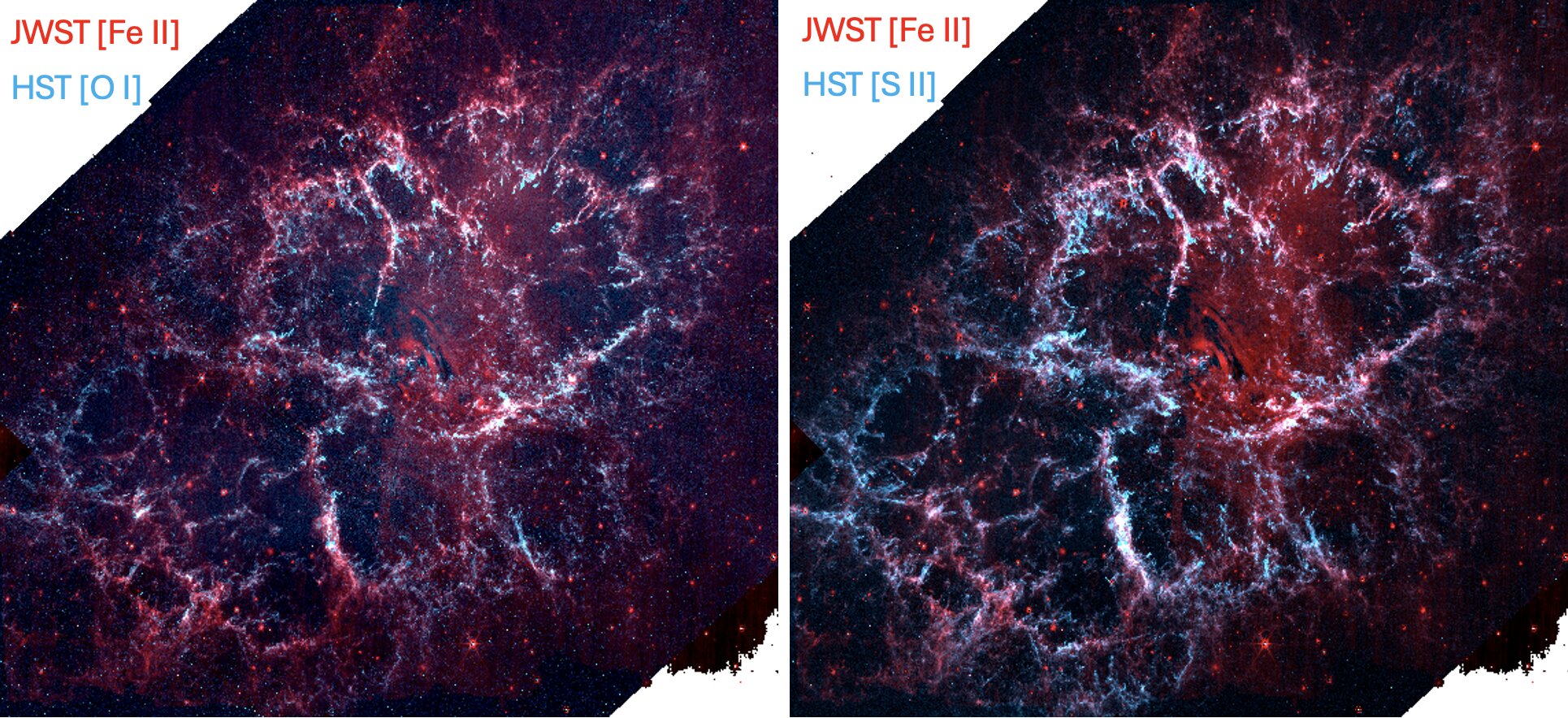}
\caption{This figure shows the JWST NIRCam \feii\ image in red (both panels), with the WFC3 \oi\ (left panel) and \sii\ (right panel) in teal. All of these ions arise from high density regions, but since they involve different elements direct interpretation of the images is difficult.
\label{comp-fe2}
}
\end{figure*}

NIRCam contains a filter that is centered on \feii\ 1.64 $\mu$m (F162M) but is also broad enough to pass significant synchrotron emission. \citet{temim24} used their F480W image to scale and subtract the continuum, leaving a fairly clean \feii\ image (with some residual synchrotron emission from the wisps in the central region near the pulsar as well as low-level diffuse emission to the NW of the pulsar). In Fig. \ref{comp-fe2}, we show this \feii\ image against the WFC3 \oi\ (left panel) and \sii\ (right panel).

Significant variations are seen in the relative intensities with position in these maps. Each ion is from a different element, so we cannot unequivocally separate ionization/density variation from compositional changes without detailed spectroscopy, kinematic information, and modeling.  However, all of these ions have low ionization potentials (7.90 eV for \feii, and 10.4 eV for \sii, while neutral O only exists below 13.6 eV). Hence, all of these sample the dense filament cores at some level, and much of the variation may be due to abundance variations.
Fig. \ref{fe2s2sections} in the Appendix shows enlargements of sections from the right panel of Fig. \ref{comp-fe2} to show detail.

\subsubsection{A Closer Look: MRS Positions \label{sec:HST-JW-MRS}}

Our program was partially motivated by the desire to compare optical imagery to selected images from JWST NIRCam and MIRI, obtained as part of Prop. 1714 (PI Temim; see \citet{temim24}).  In particular, we take advantage of HST's imaging resolution to inspect the optical emission at each of the two MRS aperture locations discussed by \citet{temim24}.

\begin{figure*}
\center
\includegraphics[width=0.8\textwidth]{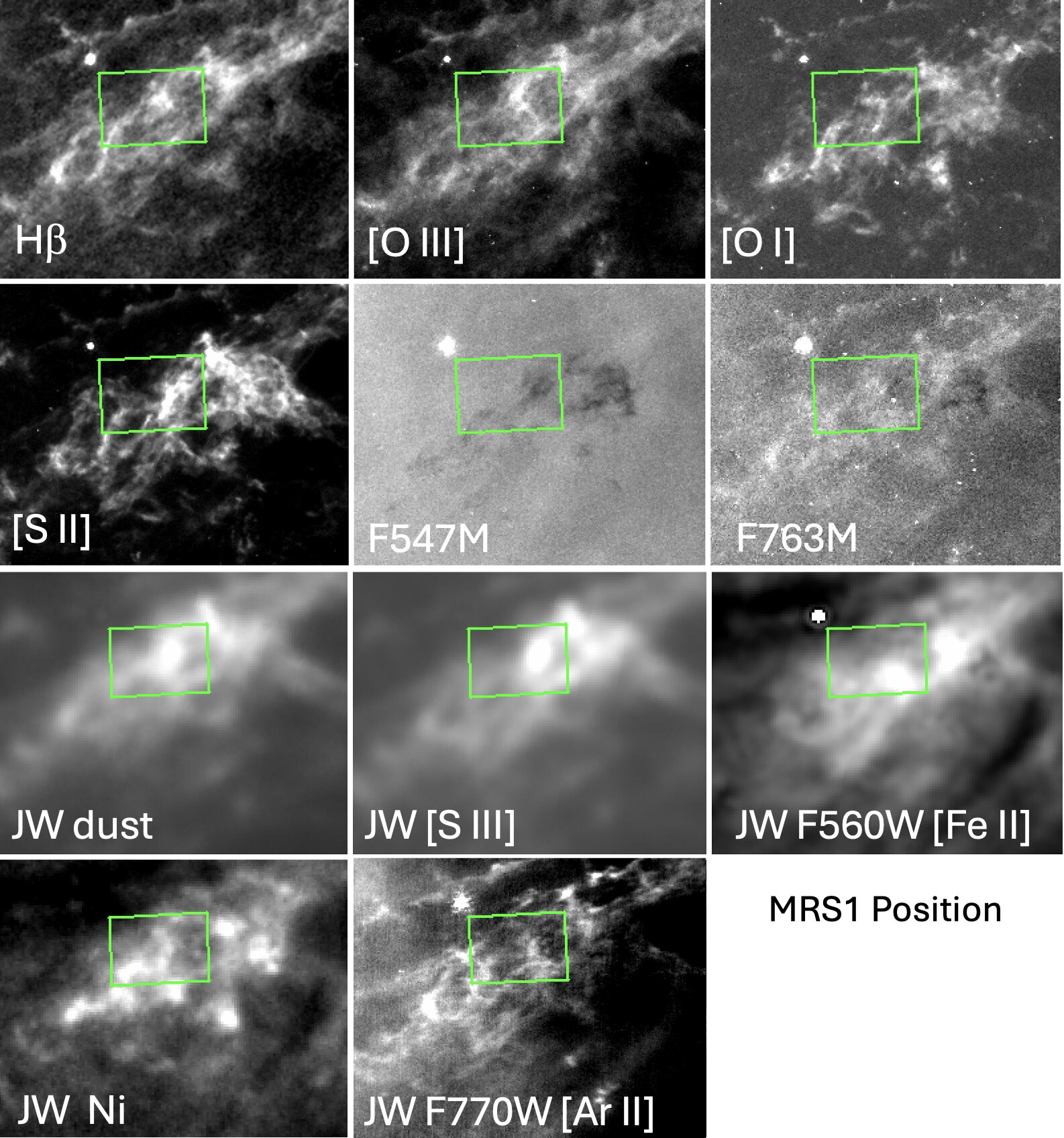}
\caption{This figure compares cutouts for the WFC3 and JWST imaging data at the MRS1 position on the central bar filaments demonstrating the complexity of the emission covered by the MRS observation.  The green box shown is 2.8\arcsec\ $\times$ 4.0\arcsec\ in size.
\label{MRS1all}
}
\end{figure*}

\begin{figure*}
\center
\includegraphics[width=0.8\textwidth]{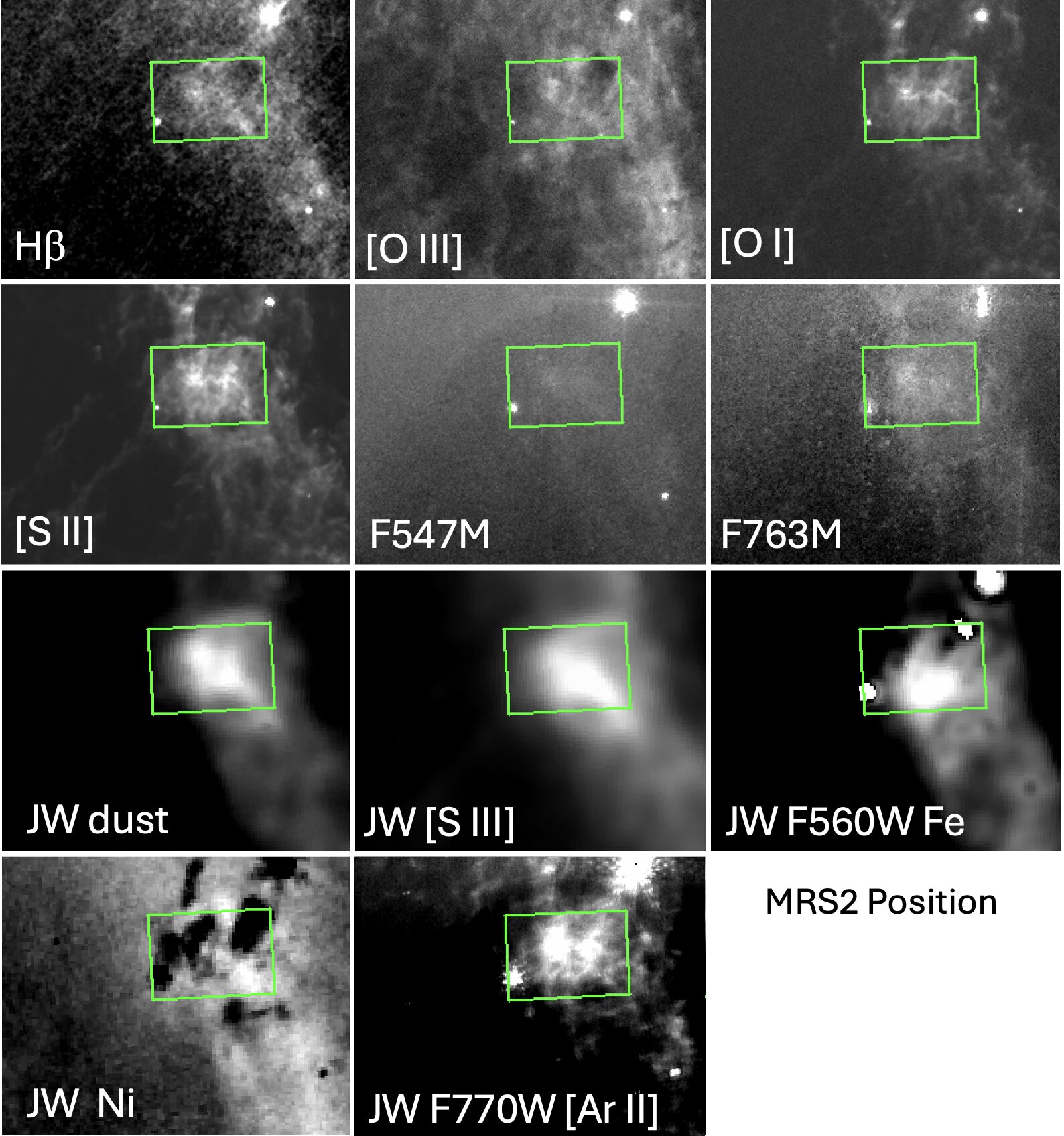}
\caption{Same as Fig.~\ref{MRS1all} but for the JWST MRS2 position. The JWST nickel map suffers from several significant susbtraction residuals from the attempt to derive the faint emission component due to \nic.
\label{MRS2all}
}
\end{figure*}

Fig.~\ref{MRS1all} and Fig.~\ref{MRS2all} are multipanel figures showing aligned regions surrounding the MRS1 and MRS2 positions, respectively, as seen with WFC3 and with the JWST cameras. The MRS1 position was centered on the central bar filaments and MRS2 was positioned on the bright southern filament.  The six available WFC3 images are shown along with five from JWST.  While JWST imagers do not in general have narrow band filters, certain filters contain sufficiently bright emission lines that some information can be derived, as discussed at length by \citet{temim24}. Also, using information from overlapping filters, the underlying dust continuum component in JWST imagery could be extracted into an image.  Finally, a ``nickel map" was constructed via fainter lines of several ionization stages of this element\footnote{The JWST/MIRI F1130W filter contains lines of [Ni~II] 10.682 $\mu$m, [Ni~III] 11.002 $\mu$m, and [Ni~IV] 11.726$\mu$m, but [Ni~II] is likely the strongest. Nonetheless, it is difficult to know how to interpret the `nickel' image.} with the bandpass of one JWST filter, F1130W.

\begin{figure*}[t]
\center
\includegraphics[width=0.5\textwidth]{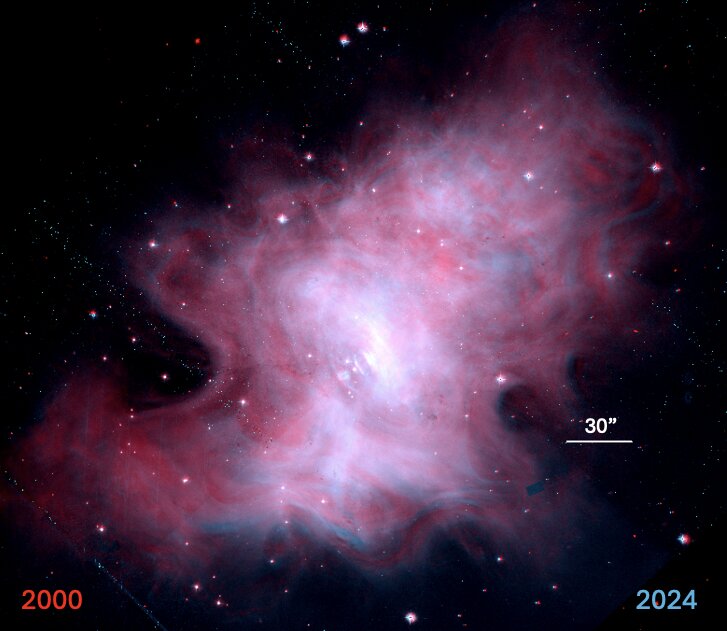}
\caption{This figure compares the large scale changes in distribution of the synchrotron nebula over the two epochs (red: WFPC2, 2000; cyan: WFC3 2024). Smaller scale banding may arise from expansion, but other more patchy color differences may imply changes in the relative intensity over the two epochs.
\label{sync_overview}
}
\end{figure*}

\begin{figure*}[b]
\center
\includegraphics[width=0.5\textwidth]{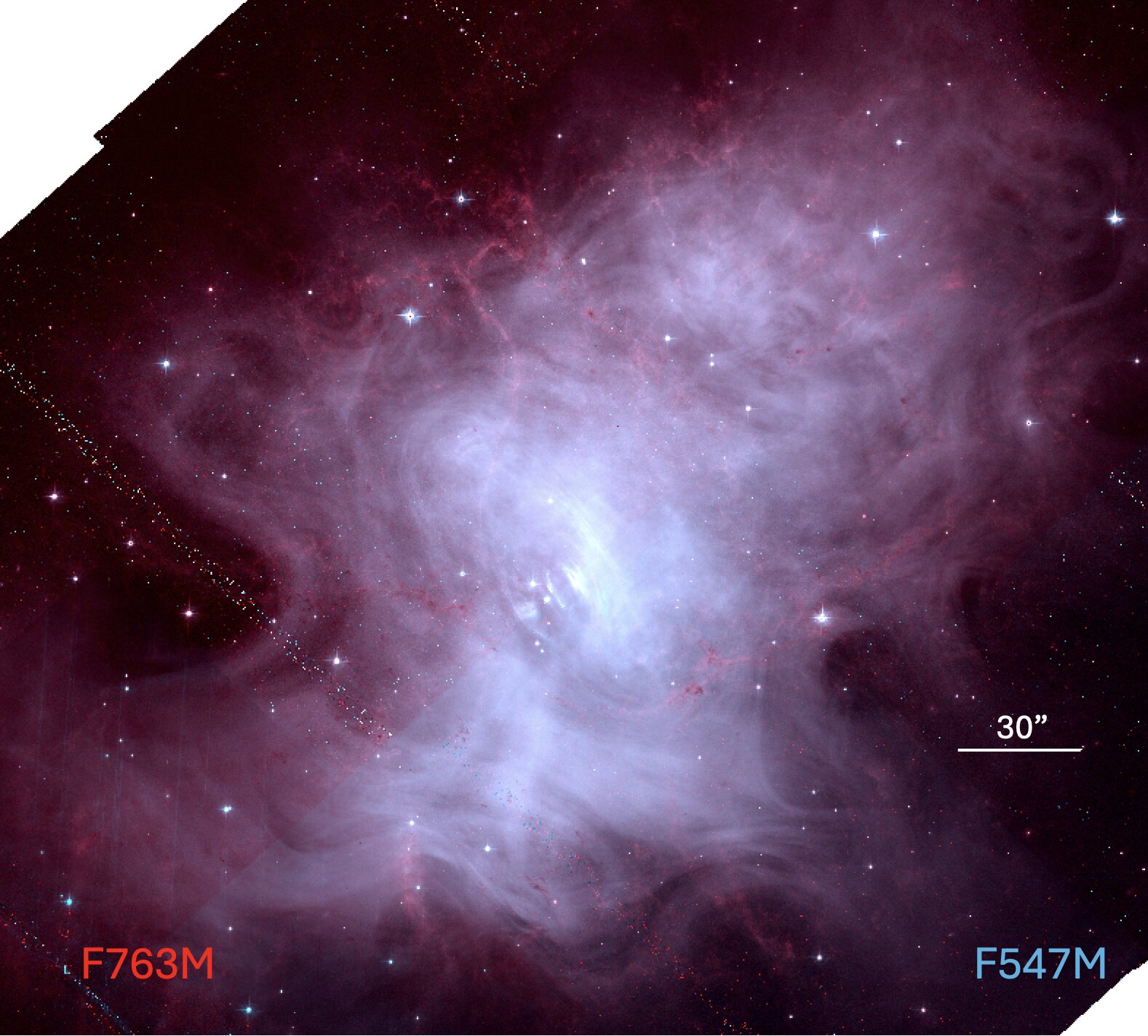}
\caption{This figure compares the synchrotron continuum emission for the Epoch 2 WFC3 data, with F547M shown in cyan and F763M in red. The flat white appearance of the nebula indicates little change in relative intensity of the synchrotron emission within the HST range, although there is a hint of redder coloring toward the outside edges of the nebula.  We note that F763M also passes some emission from filaments, primarily \nicL, which is known to be strong and variable with position.
\label{cont-comp}
}
\end{figure*}

\begin{figure*}[t]
\center
\includegraphics[width=0.6\textwidth]{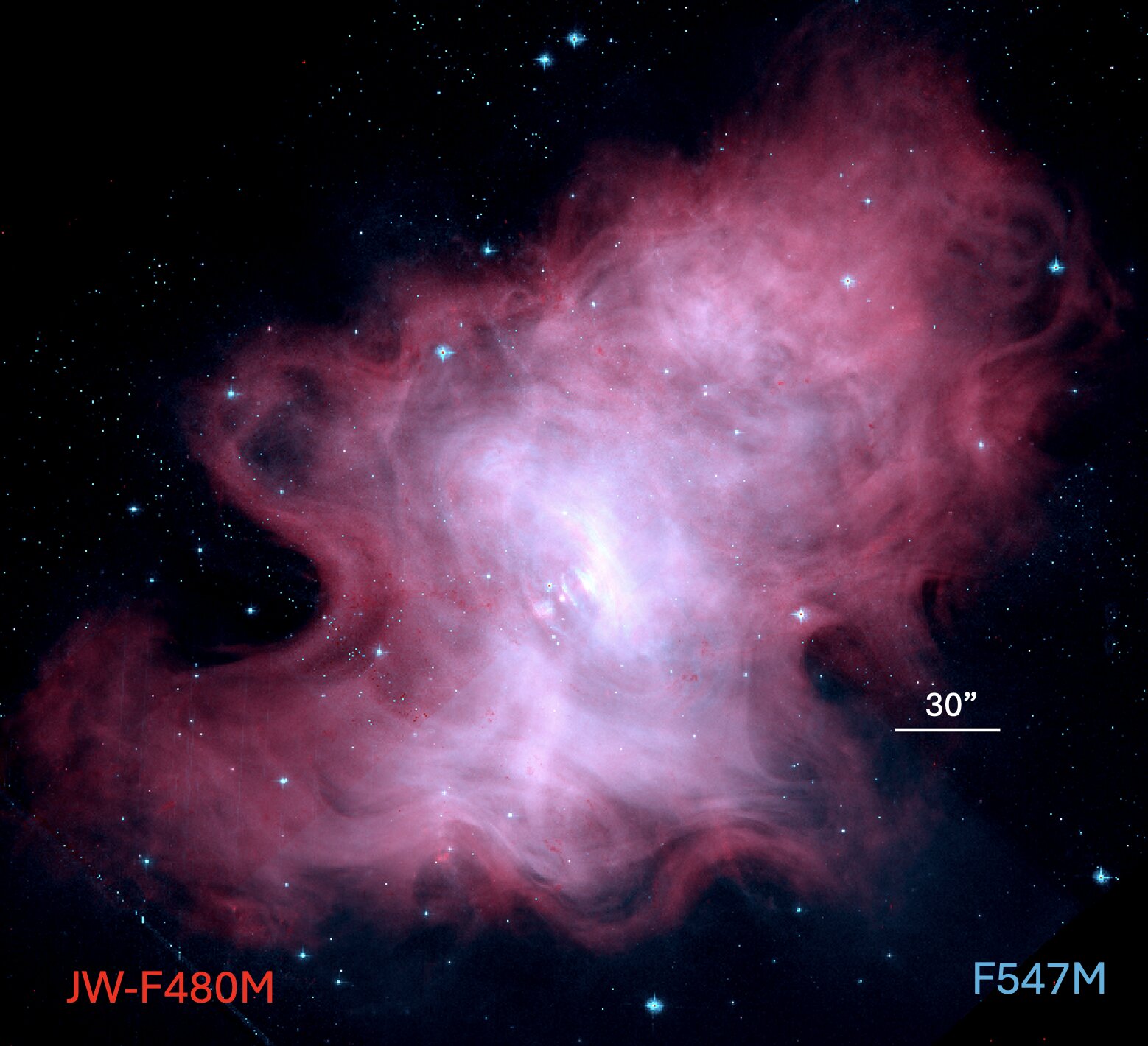}
\caption{This figure compares the appearance of the synchrotron nebula at JWST 4.8 $\mu$m (NIRCam F480M, red) and optical (WFC3 F547M, cyan), providing a longer baseline.  Here the redder coloring toward the outer parts of the nebula appears more obvious. These two data sets were only obtained roughly 14 months apart, so nebular expansion should be minimal.
\label{jwst-comp}
}
\end{figure*}

Inspection of these images shows the complexity of the emission at each of these locations in the Crab, including within the MRS apertures themselves where the positions and morphologies of the brightest filaments appear different in each emission line. Certainly part of this variation is due to differing kinematic components, as noted by \citet{temim24}, but changing composition and ionization are also likely occurring.   Furthermore, the resolution varies somewhat between WFC3 and JWST/MIRI.  Most structure seems to be resolved by the 0.04\arcsec\ pixels in the WFC3 images, but the MIRI imaging resolution changes with wavelength (see Table 1 of \citealt{temim24} for MIRI resolution versus wavelength).  The shorter wavelength MIRI \feii\ and \arii\ images show only slighter worse resolution of $\sim$0.20--0.25\arcsec, while the MIRI \siii\ and dust maps (both derived from longer wavelength MIRI images are noticibly lower resolution ($\sim$0.6\arcsec). The nickel image, derived from the MIRI F1130W filter, is intermediate in resolution between these. Despite these changes, most of the relevant structures are visible from image to image.

In Fig.~\ref{MRS1all} for the MRS1 position on the central bar region, both similarities and differences are readily seen betwen these various images.  The \sii\ and \siii\ distributions are similar, with some correlation to structure seen in \oiii\ as well. As seen earlier, the dust emission map from JWST does not match the dust shadows seen in F547M in detail, although they might be measuring different aspects of the dust.  The bright \feii\ knots do not align with features in the other bands. Interestingly, the bright knots and overall appearance of the JWST nickel image agrees better with WFC3 \oi\ than any of the other images. 

At MRS2 (Fig.~\ref{MRS2all}), the two-knot structure seen in \arii\ appears to correlate well with \sii\ and \oi, indicating spatial agreement for these low ionization emissions.  The bright knot in \feii\ aligns fairly well with bright \siii\ in this case, but the morphology of these two emission do not agree in detail. There is a lack of correlation seen between \siii\ and \oiii\ in conparison to MRS1, perhaps indicative of a more significant abundance variation. \citet{temim24} indicate somewhat red-shifted emission from this filament region, so the absence of a dust shadow to correlate with the JWST dust emission is not surprising. Certainly more detailed modeling of these filament locations is needed to separate possible abundance and ionization/density variations that may be present.

\subsection{Synchrotron Comparisons \label{sec:HST-JW-Sync}}

The interior synchrotron nebula close to the pulsar is known to be dynamic on short time scales \citep{hester02}.  However, on a longer time scale, one might expect changes in the more extended nebular structure as well, as energy leaks away from the local influence of the pulsar.  The long time baseline now available and the similarities of the F547M filters from WFPC2 and WFC3 make this possible to investigate.

Fig.~\ref{sync_overview} shows a two-color comparison of the WFPC2 and WFC3 data sets, with Epoch 1 in red and Epoch 2 in cyan.  If no changes had occurred, the diffuse nebula would simply be uniformly white, but there are clearly patchy regions of red and cyan in the figure showing that there have indeed been large scale changes over the ensuing 24 years.

In Epoch 2, we obtained two medium bands with WFC3 dominated by synchrotron emission (albeit with some minor contamination from emission filaments).  In Fig.~\ref{cont-comp}, we show the WFC3 F547M and F763M data in two-color format.  The lack of patchy color variations in this figure indicates that any variations of the spectral index between these two bands are minor. There is some indication that the outer fringes of the nebula are tinged toward the red, indicating F763M is stronger there.

The JWST NIRCam F480M image is also dominated by synchrotron continuum and provides a longer baseline.  Fig.~\ref{jwst-comp} shows a comparison of F480M (red) to WFC3 F547M (cyan).  The figure shows weaker optical wavelength emission in the outer regions of the nebula compared with the infrared image.  This is likely associated with expected synchrotron cooling (see
\cite{lyutikov19, dirson23}).  The more dramatic effect for the longer wavelength baseline compared with the two optical bands is consistent with this picture.


\subsection{Two Surprising Features
\label{sec:surprising}}

\begin{figure*}
\center
\includegraphics[width=0.9\textwidth]{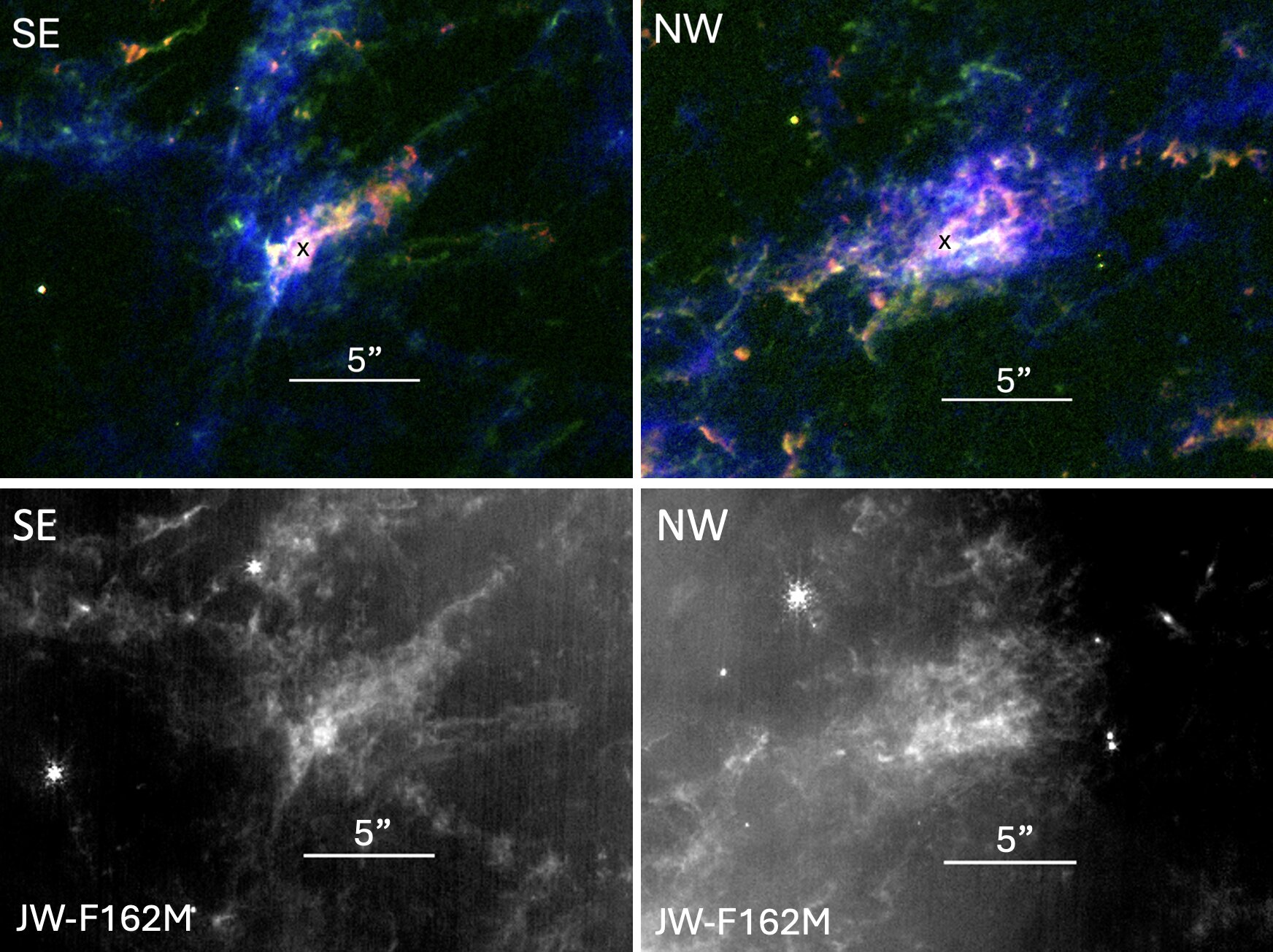}
\caption{The top two panels in this figure show the subtracted 3-color WFC3 emission line data for the two previously unrecognized regions of enhanced emssion that are almost diametrically opposed to the pulsar position (see white boxes in Fig.~\ref{subtracted_color}.  The black `x' in each panel marks the fiducial position used to assess position and motion with respect to the pulsar, as described in the text.  The bottom two panels show the data from the JWST NIRCam \FeiiL\ (F162M) image for the same regions.
\label{new_features}
}
\end{figure*}

Inspection of the subtracted WFC3 emission line images  in Fig.~\ref{subtracted_color} reveals two compact regions of filaments that stand out from their surroundings, as highlighted by the white boxes.  These filament groupings are nearly diametrically opposed to the pulsar but are at large separation. The top two panels of Fig.~\ref{new_features} show close-up views of these two regions using the same color scheme as Fig.~\ref{subtracted_color} but with intensity scaling adjusted to avoid saturation and show the detailed filament structures. These features are present in the Epoch 1 data, but they did not stand out in the figures shown by \citet{loll13}.  The NW feature is brighter than the SE feature, but both stand out from surrounding emission, showing a complex morphology of clumpy knots, filaments, and more diffuse emission.   We displayed the \citet{loll13} data using the color scheme used here and find that both features show little or no change in morphology (within the limitations of slightly different spatial resolution between the data sets).  Much of the fainter emission surrounding the clumps of bright filaments arises due to unrelated material along each line-of-sight through the overall nebula.

\begin{figure*}
\center
\includegraphics[width=0.7\textwidth]{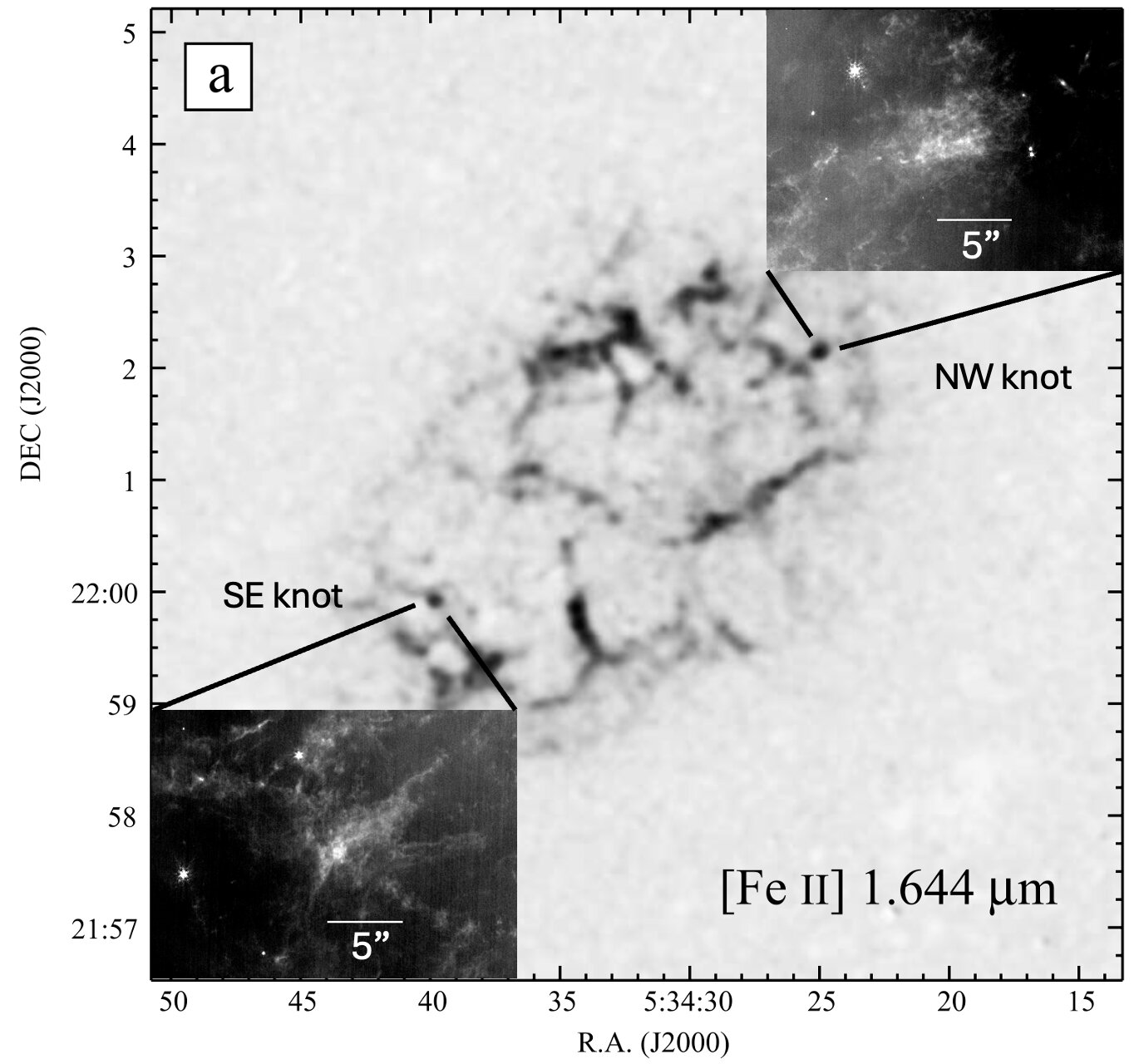}
\caption{The background is Fig. 11a from \citet{temim06}, showing a continuum-subtracted ground based \FeiiL\ image. The insets show the JWST NIRCam F162M image sections from Fig. ~\ref{new_features}, showing the detailed structure in \feii. 
\label{temim_fe2}
}
\end{figure*}

To our knowledge, these features have not been called out by previous investigators. Their positions nearly diametrically opposed to the pulsar suggest a possible association. \citet{dubner17} show various multiwavelength data sets of the Crab. The NW feature can be seen in VLA data \citep[see also][]{bietenholz90}, but the SE feature is just barely detected.  This situation is quite similar with Spitzer, where the NW feature is clearly present in IRAC 5.6 $\mu$m, 8 $\mu$m, and MIPS 24 $\mu$m (all bands containing emission lines) with a smaller enhancement at the SE position \citep{temim06}. The JWST \siii\ and dust maps \citep[][see Fig. 10]{temim24} also show the features plainly, although again, not at a level that made them stand out. Figure 7 of \citet{loll13} indicates the features are slightly offset from the major axis of the Crab, with the NW feature slightly below and SE feature above this axis.

To characterize these features, we measured their positions relative to the pulsar and their motion between the two epochs.  Since the features are extended and complex, we defined a fiducial position on a bright knot in each, as shown by the X's in Fig.~\ref{new_features}.  In Epoch 2, we find the SE feature to be 127.9\arcsec\ away from the pulsar and at an angle of $27.9^{\circ}$ below an east-west line, whereas the NW feature is 125.1\arcsec\ away from the pulsar along a line that is $38.1^{\circ}$ above the east-west line.  The motion between the two epochs is directed along these same angles, with the NW feature moving away from the pulsar by 3.7\arcsec\ $\pm$ 0.1\arcsec, while the SE feature moved by 2.8\arcsec\ $\pm$ 0.1\arcsec\ in the opposing direction.


We have also used the previously mentioned SITELLE data to investigate the kinematics of these features.  We concentrate on the velocities at \oiii\ where the component structure is simpler to interpret.  While both regions show a range of velocities, the NW clump is dominated by a peak at 68 $\kms$ with a second weaker component at roughly -260 $\kms$. Hence, this clump is moving nearly in the plane of the sky. The SE clump, however, shows at least two main peaks, both at significantly negative velocities (-590 and -840 $\kms$).  Hence, the SE clump is moving significantly out of the plane of the sky and toward the observer.  It appears the near symmetry seen in position relative to the pulsar is not mirrored in their velocity structure.

The bottom two panels of Fig.~\ref{new_features} show the same regions as the upper panels, but as they appear in the F162M \FeiiL\ image from JWST.  Once again, the filaments are seen to have significant detailed structure that makes them stand out from their surroundings. However, the brightness of the features in \feii\ is more obvious in the ground-based \feii\ image reproduced in Fig. ~\ref{temim_fe2}, adapted from Fig. 11a of \citet{temim06}.  Both features stand out rather well in \feii.  In other settings, \feii\ is well established as an indicator of shock heating, especially given that shocks with velocity greater than $\sim$100$\kms$ can destroy dust, thus elevating gas phase Fe abundances \citep[cf.][]{koo16}.  This may or may not be the case here since the knots and filaments in the Crab are dominated by low ionization, dense cores.  In general, the Crab's filaments are photoionized by the hard synchrotron spectrum due to the pulsar which can affect relative line strengths. There do not appear to be local enhancements in the synchrotron emission near the positions of these features, however, so shock effects cannot be discounted as a possible excitation mechanism for these features.   
Further study of these features is needed to understand whether they are related to possible activity by the pulsar at some time in the past or some other cause.

\section{Summary and Conclusions}

It has been nearly a quarter of a century since the full extent of the Crab Nebula has been imaged with HST.  The WFC3 filters used for emission line imaging reported here are similar to those from Epoch 1 (observed using WFPC2 in 1999--2000) but capture substantially more of the blueshifted emission.  HST imaging apparently resolves the majority of the ejecta knots and filaments, with relatively few knots appearing to be at point source resolution.

Comparison of data from the two epochs shows the variable proper motion of filaments with position, as expected from the 3D expanding nebula. Filaments toward the outer regions of the nebula show the highest motion in the plane of the sky while much less proper motion is seen within the inner regions. The Rayleigh-Taylor filaments reported in earlier data appear to have largely participated in this expansion and show little or no stretching or variation other than expansion. Variations in the large scale structure of the synchrotron continuum are also readily visible over the 24 year separation. These new high resolution images will be the most applicable for comparison to contemporary multiwavelength data sets for the foreseeable future.

The ejecta knots and filaments show dramatic spatial differences in the three primary emission lines sampled with WFC3 filters.  Much of this variation is likely attributable to density variations which in turn are affecting the observed ionization structures.  While compositional variations may also play a role, this can only be assessed with detailed spectroscopy and modeling, including also an assessment of the kinematics of individual filaments, now available from SITELLE 
\citep[][Ding et al., in preparation]{martin21,martin25}. 
As such, the new data presented here become the springboard for ongoing detailed studies of this dynamic object, including comparisons with contemporaneous multiwavelength data sets.

\vspace{0.3in}

This work has been supported by STScI grant HST-GO-17500-A (WPB) and subordinate grants under program 17500 to the coauthors.
WPB acknowledges support from the Dean of the Krieger School of Arts and Sciences and the Center for Astrophysical Sciences at JHU during this work. JML was supported by basic research funds of the Office of Naval Research. DM acknowledges NSF support from grants PHY-2209451 and AST-2206532.

\facilities {HST (WFPC2, WFC3), CFHT (SITELLE)}

\software {SAOimage ds9, DrizzlePac 3.0}

\clearpage

\bibliographystyle{aasjournal}

\bibliography{Crab.bib}

\clearpage


\begin{deluxetable}{ccll}
\tablecaption{HST Program 17500 WFC3 \\ Observations of the Crab Nebula}

\tablehead{
 \colhead{Field ID} & 
 \colhead{Date(obs)\tablenotemark{a}} & 
 \colhead{RA(J2000)\tablenotemark{b}} & 
 \colhead{Dec(J2000)\tablenotemark{b}}  
}
\tablewidth{0pt}
\startdata
1  & 2024 Feb 22  & 83.6656258 & 21.9866916   \\
    & 2024 Apr 08\tablenotemark{c}  & 83.6656258 & 21.9866916   \\
2  & 2024 Feb 23  & 83.6247413 & 21.9907777   \\
    & 2024 Apr 08\tablenotemark{c}  & 83.6247413 & 21.9907777   \\
3  & 2024 Feb 25  & 83.5887021 & 22.0207167   \\
4  & 2024 Feb 27  & 83.6603054 & 22.0190805   \\
5  & 2024 Mar 01  & 83.6027600 & 22.0442167   \\
6  & 2024 Mar 06  & 83.6299583 & 22.0452361   \\
H$\beta$(N)  & 2024 Feb 15  & 83.6217879 & 22.0586944   \\
H$\beta$(S)  & 2024 Feb 17  & 83.6656258 & 21.9937083   \\
\enddata
\tablenotetext{a} {All data obtained at V3PA = 267 degrees.}
\tablenotetext{b} {Position of UVIS-CENTER reference point.}
\tablenotetext{c} {Re-observations of F547M and/or F763M due to guide star loss-of-lock on initial observations.}
\label{fields}
\end{deluxetable}

\begin{deluxetable}{cccc}
\tablecaption{Crab Nebula HST Program 17500 WFC3 Exposure Information\tablenotemark{a} }
\tablehead{
 \colhead{Filter} & 
 \colhead{N(exp)\tablenotemark{b}} & 
 \colhead{PostFlash} & 
 \colhead{$\rm t_{exp}$(Total) (s)} 
}
\tablewidth{0pt}
\startdata
F502N  & 3 & 20 & 2460   \\
F547M  & 2  & 10 & 1262   \\
F631N  & 3 & 20 & 3983   \\
F673N  & 3  & 15 & 3797   \\
F763M  & 2  & 15 & 1062   \\
F487N  & 4\tablenotemark{c}  & 20 & 7808   \\
\enddata
\tablenotetext{a} {The exposure information for the first five filters listed was repeated for each of the six mosaic fields.}
\tablenotetext{b} {Three exposure pattern used WFC3-DITHER-LINE-3PT with 2.414\arcsec\ steps at 85.759 degrees.  Two exposure pattern used WFC3-DITHER-LINE with 2.414\arcsec\ steps at 85.759 degrees. }
\tablenotetext{c} {H$\beta$ observations were obtained for two central fields and do not correspond directly with any of the six fields used for the full mosaics; see text.  F487N used a two exposure pattern WFC3-DITHER-LINE with 2.414\arcsec\ steps at 85.759 degrees but was repeated over two orbits to get the total listed exposure time.}
\label{filters}
\end{deluxetable}

\begin{deluxetable}{lcrrrrrrr}
\label{comparison}
\tablecaption{WFPC2 and WFC3 Filter Comparison}
\tablehead{
\colhead{Filter} &
\colhead{Ion} &
\colhead{Width} &
\colhead{$\lambda$(blue)} &
\colhead{$\lambda$(red)} &
\colhead{$\lambda$(ref)} &
\colhead{$\rm V_{low}$} &
\colhead{$\rm V_{high}$}
\\
\colhead{} &
\colhead{} &
\colhead{(\AA)} &
\colhead{(\AA)} &
\colhead{(\AA)} &
\colhead{(\AA)} &
\colhead{($\rm km ~s^{-1}$)} &
\colhead{($\rm km ~s^{-1}$)}
}
\startdata
\\
WFPC2: & & & & & & & &  \\
F502N & [O~III]	& 27	& 4999 &  5026 & 5007 & $-$480 & 	1140  \\
F631N &  [O~I]	& 30	& 6291 &  6321 & 6300 & $-$429 & 	1000  \\
F673N & [S~II]	& 45	& 6709 &  6756 & 6725 & $-$560 & 	1740  \\
F547M &	cont. & 650	& 5122	& 5772 &  -   &   -  & -  \\ 
\\
WFC3: & & & & & & & &  \\
F487N & H$\beta$ & 60	& 4841 & 4901 &	4861 & 	$-$1234 &	2419  \\
F502N & [O~III]	& 65	& 4978 & 5043 &	5007 & 	$-$1738 &	2157  \\
F631N &  [O~I]	& 58	& 6275 & 6333 &	6300 & 	$-$1190 &	1571  \\
F673N &  [S~II]	& 118	& 6707 & 6827 &	6725 & 	$-$960 &	$>$2500  \\
F547M &	cont. & 650	& 5122	& 5772  &  -   &   - &  -  \\  
F763M &	cont. & 704	& 7250	& 7962  & -    &  -  &  -  
\enddata
\end{deluxetable}


\clearpage

\appendix

\setcounter{figure}{0}
\renewcommand\thefigure{A\arabic{figure}}

In this Appendix, we highlight the incredible complexity of the Crab Nebula's ejecta filaments using color figures generated from the three primary emission line data sets obtained with WFC3. With 0.04\arcsec\ pixels, the WFC3 data are even better resolution than the Epoch 1 WFPC2 data. At 2.0 kpc, a WFC3 UVIS pixel corresponds to $1.20 ~ \times\ 10^{15}$ cm, or $3.9 ~\times\ 10^{-4}$ pc. As noted by \citet{blair97}, the vast majority of filamentary structure in the Crab appears to be resolved at these levels.

For the three-color figures below, we use the same color scheme and linear stretch used in the continuum-subtracted  Fig.~\ref{subtracted_color} (\oiii\ in blue, \oi\ in green, and \sii\ in red).  Regions bright in all three bands appear white, while those bright in \sii\ and \oi\ appear yellow, and other combinations provide a range of colorations.

\begin{figure}[b]
\center
\includegraphics[width=1.0\textwidth]{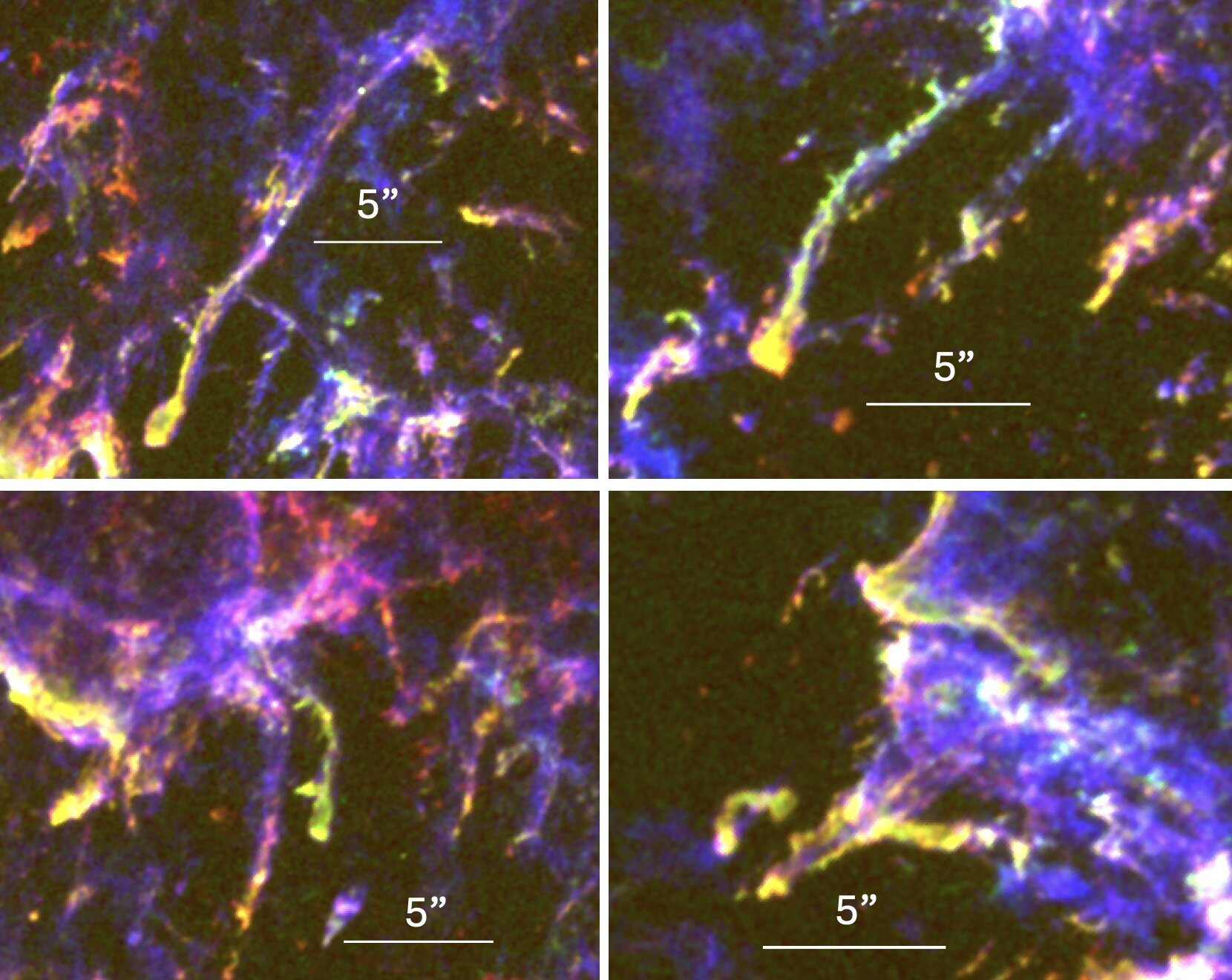}
\caption{This four panel figure shows regions dominated by magnetic Rayleigh-Taylor finger filaments. These features are dominated by low ionization, and hence appear bright in \sii\ and \oi\ (thus mainly appearing yellow and green).
\label{A1fig}
}
\end{figure}

\begin{figure}
\center
\includegraphics[width=1.0\textwidth]{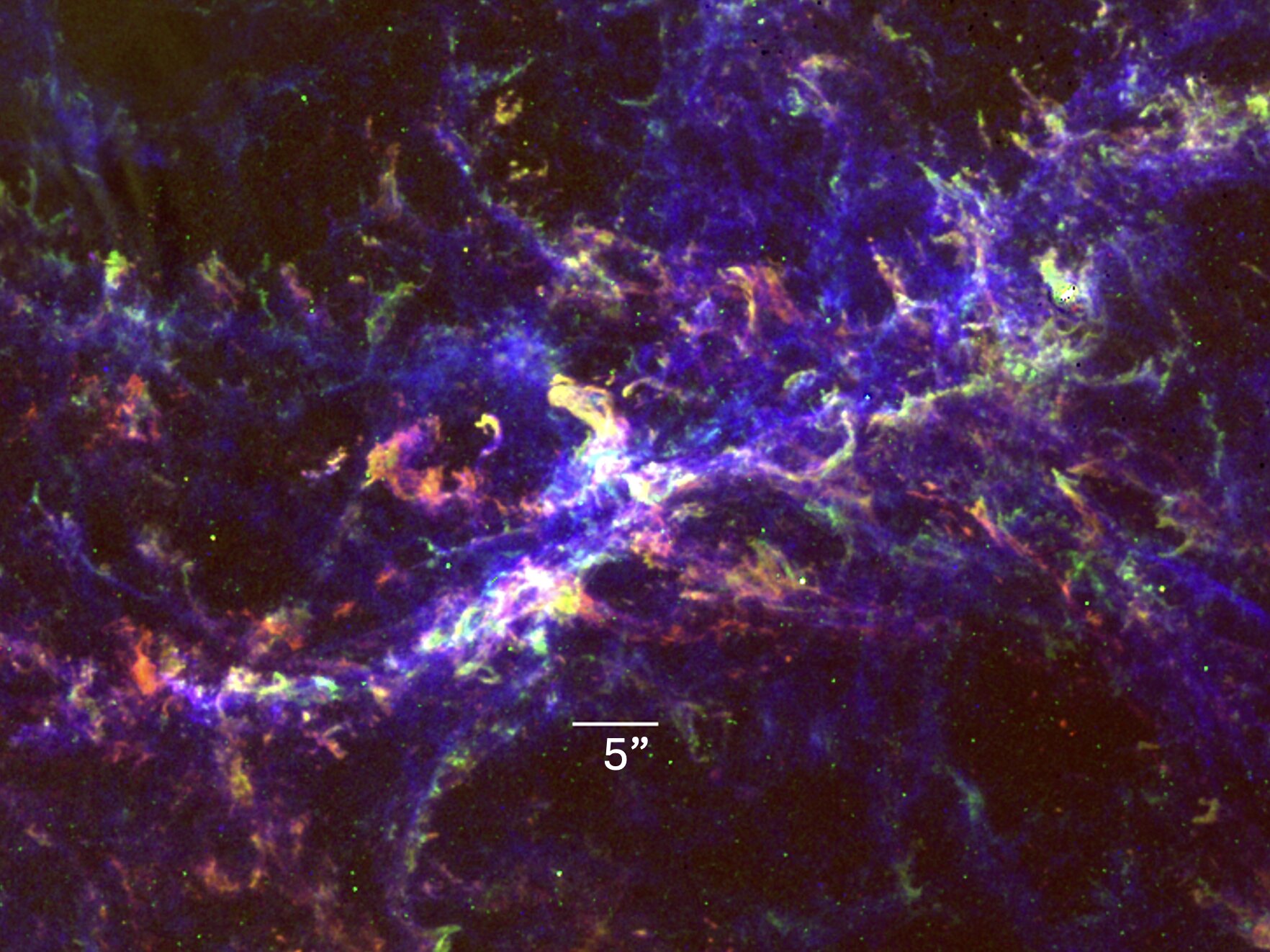}
\caption{This figure highlights the central bar region and includes the JWST MRS1 position (cf. \cite{temim24}). The compact dense cores of filaments in this region are embedded within more extended and diffuse higher ionization \oiii\ emission, shown in blue.  Intensity scaling is the same as in Fig. \ref{A1fig}.
\label{A2fig}
}
\end{figure}

\begin{figure}
\center
\includegraphics[width=1.0\textwidth]{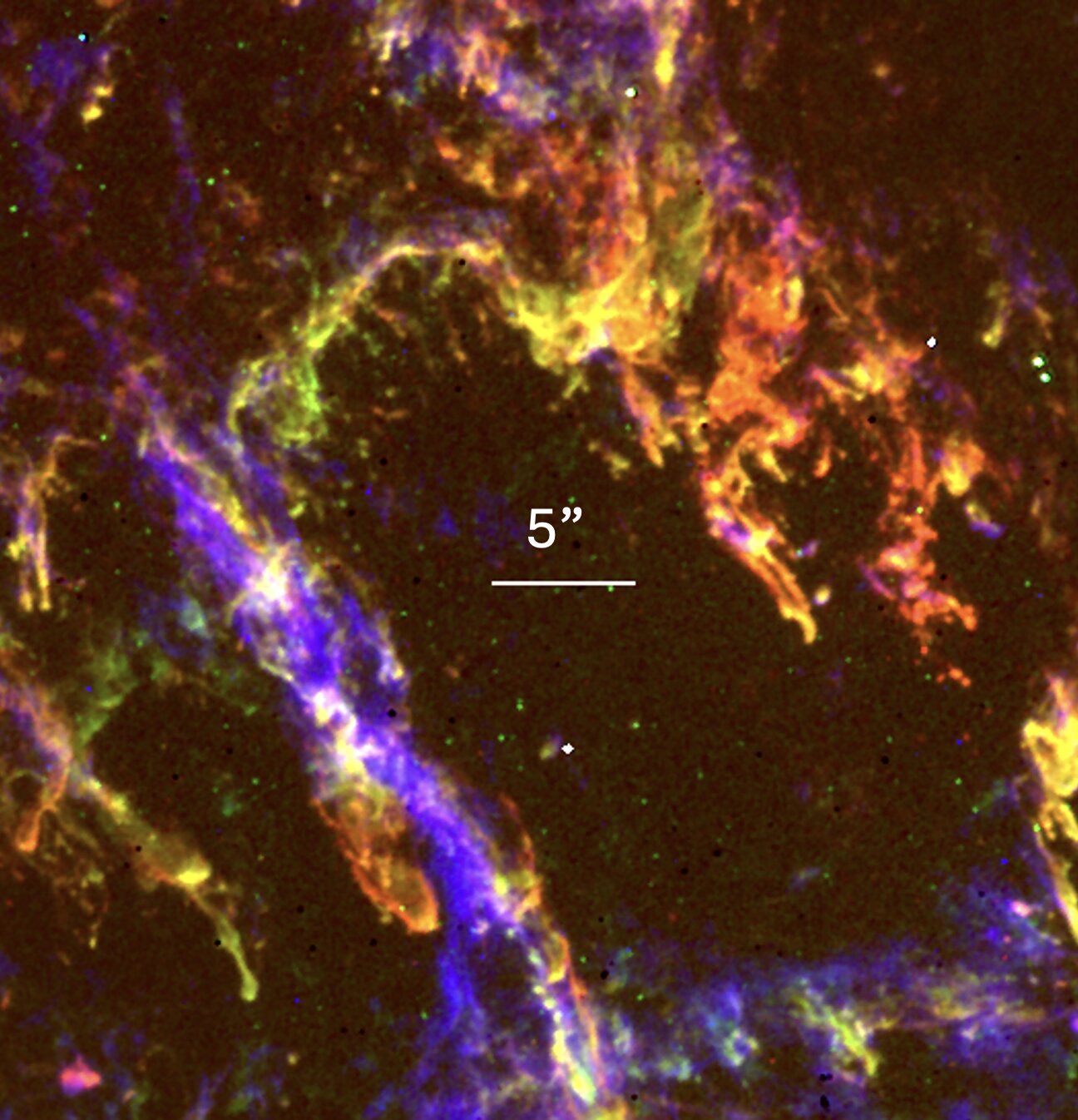}
\caption{This figure shows the so-called northern hook region.  The hook is obviously a projection of several different filament groupings with very different ionization (and hence density) properties. Intensity scaling is the same as in Fig. \ref{A1fig}.
\label{A3fig}
}
\end{figure}

\begin{figure}
\center
\includegraphics[width=1.0\textwidth]{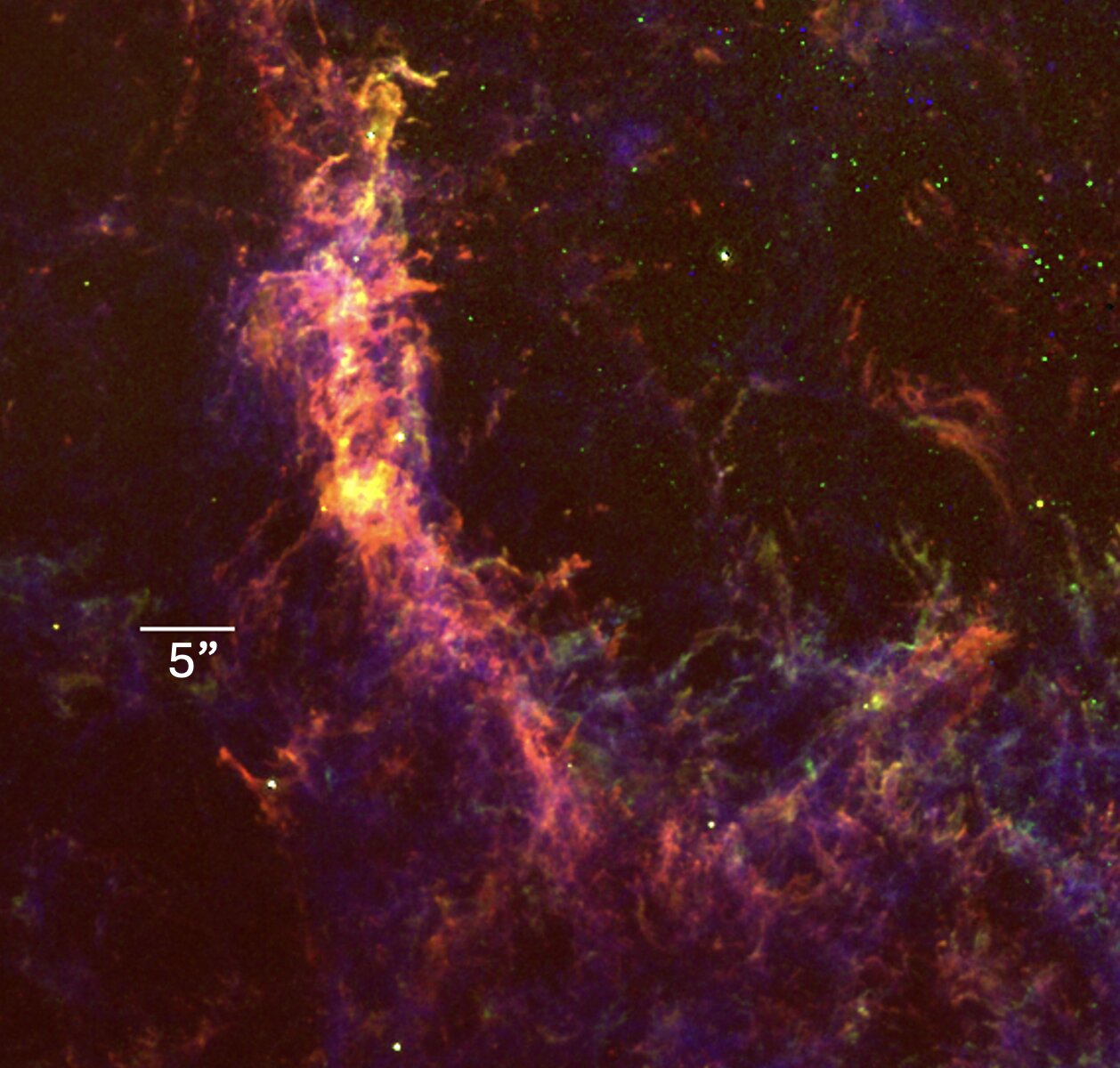}
\caption{This figure shows the bright southern filament, including the bright knot selected for the JWST MRS2 position (cf. \cite{temim24}).  This figure uses the same scaling as the previous three figures in the Appendix, but the near absence of \oiii\ (blue) and relatively weak \oi\ compared to \sii\ results in a much different (reddish) appearance.
\label{A4fig}
}
\end{figure}

\clearpage

\begin{figure}
\center
\includegraphics[width=0.95\textwidth]{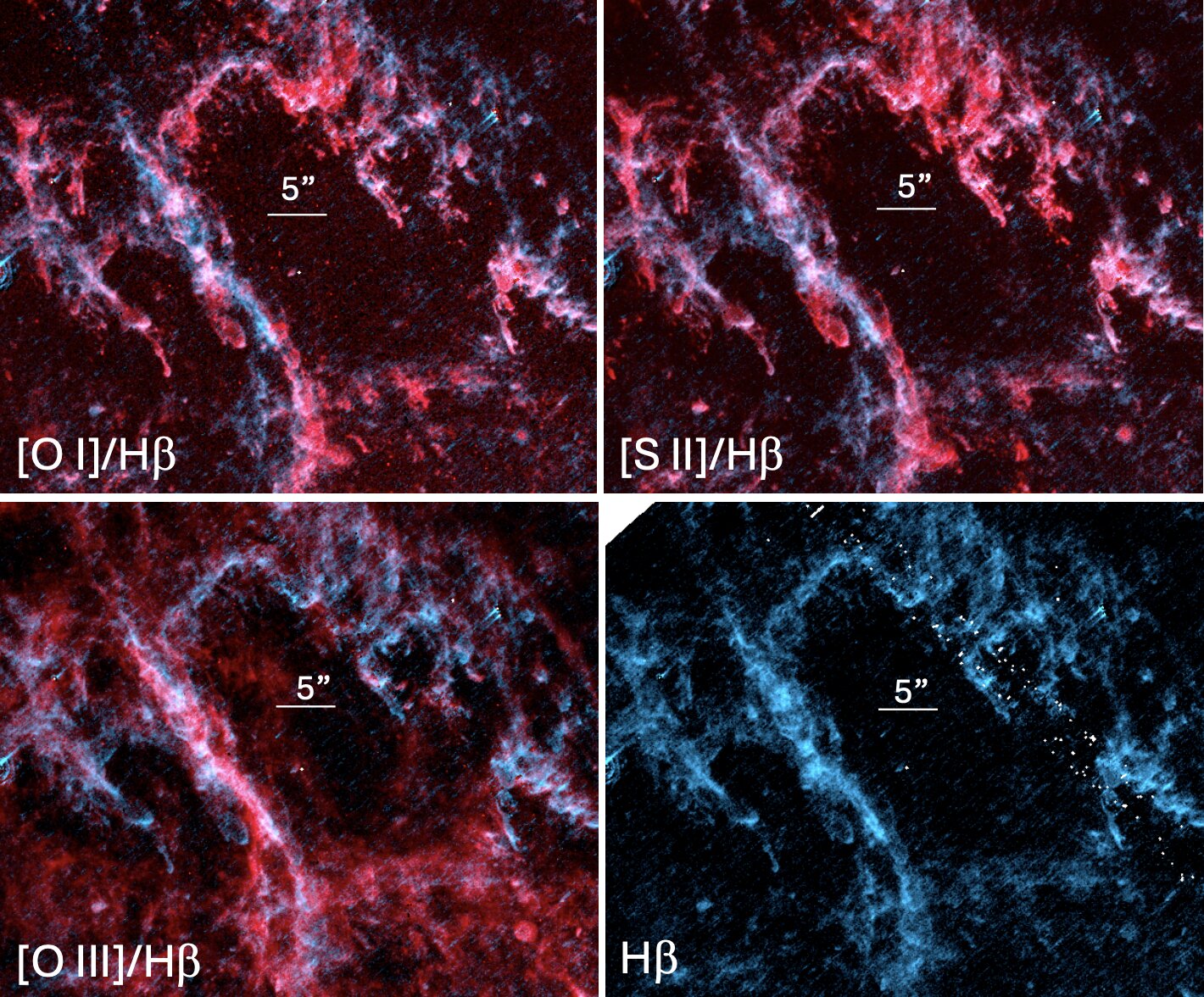}
\caption{This figure shows the bright northern hook region of filaments, in a manner similar to Fig. \ref{comp-south}, where \hb\ (in cyan) remains the same for all panels, but the indicated ions are shown in red.  The appearance and distribution of \oiii\ emission is dramatically different, but is largely due to ionization (density) differences with position. However, variations in \oi\ and \sii\ with respect to \hb\ are likely due at least in part to real compositional changes.
\label{comp-north}
}
\end{figure}

\begin{figure}
\center
\includegraphics[width=0.95\textwidth]{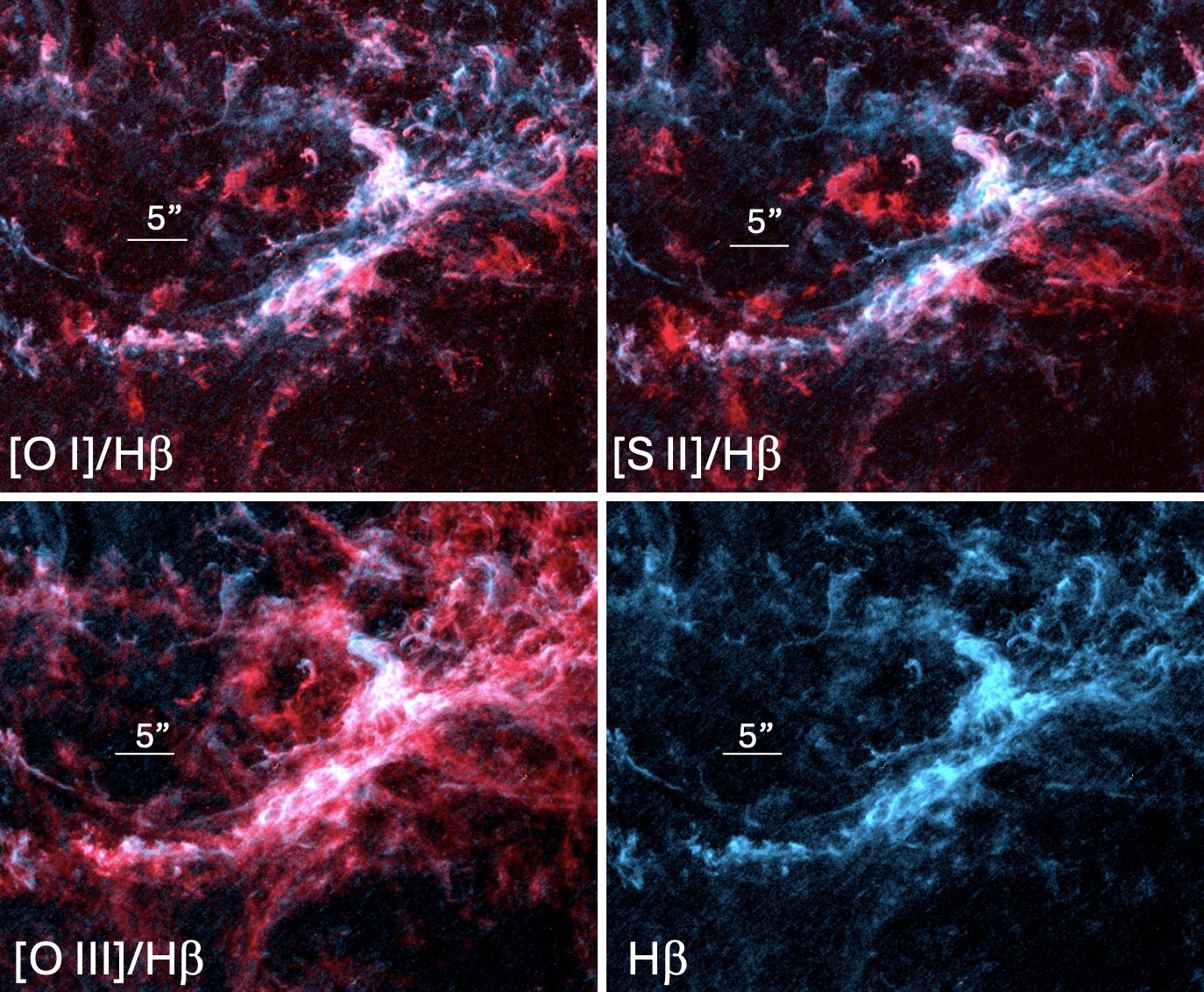}
\caption{Similar to Fig. \ref{comp-north} but for the bright central filament region.  Note how much more \oiii\ dominates in this region.
\label{comp-center}
}
\end{figure}

\begin{figure}
\center
\includegraphics[width=0.95\textwidth]{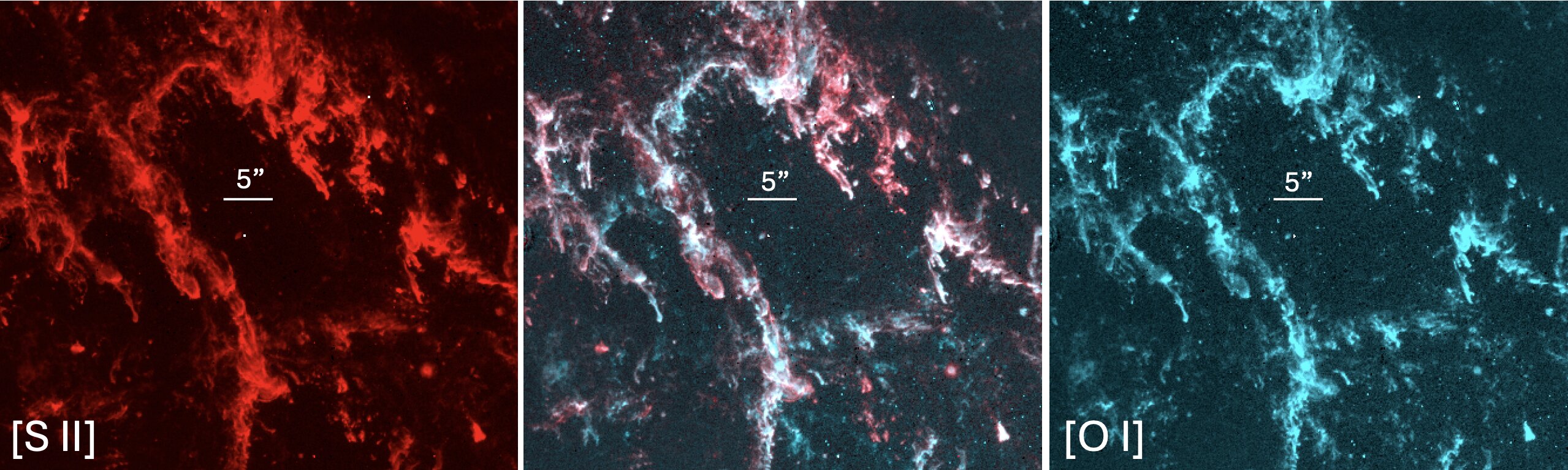}
\includegraphics[width=0.95\textwidth]{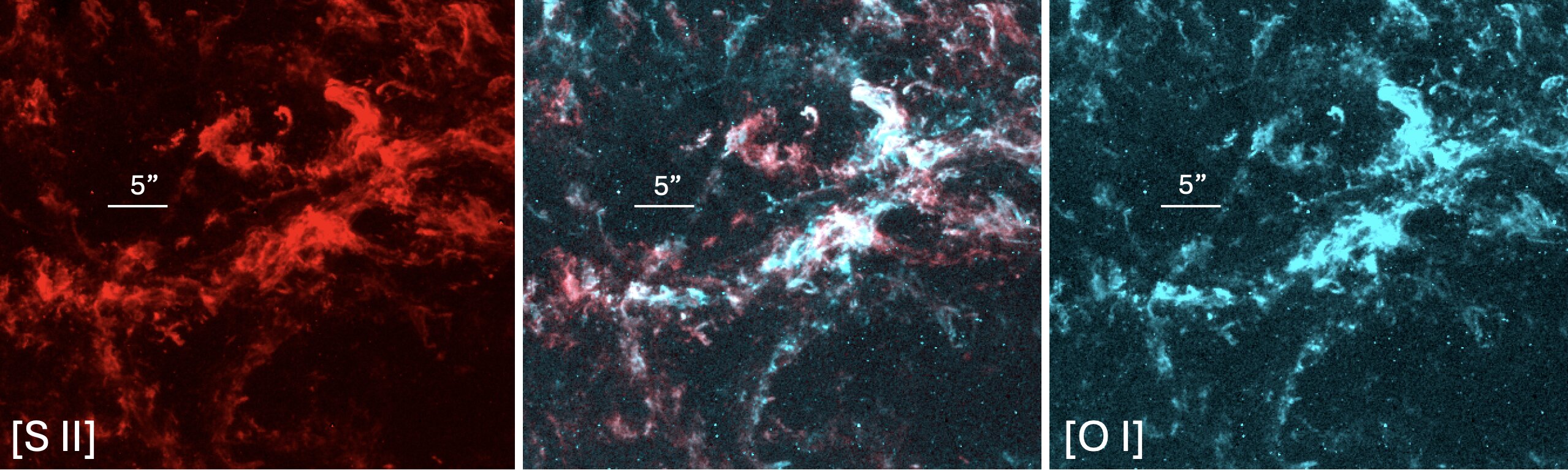}
\caption{This figure compares the appearance of the low ionization lines of \sii\ (red, left) and \oi\ (cyan, right)  individually and then combinedin the center panels to highlight the spatial variations in relative line intensities.  The top three panels show the northern hook filament region while the bottom three panels show the bright central filament region.  Since the ionization potentials of these two ions are not too disimilar, at least some of the variation seen is likely due to compositional changes.
\label{abund1}
}
\end{figure}

\begin{figure}
\center
\includegraphics[width=0.95\textwidth]{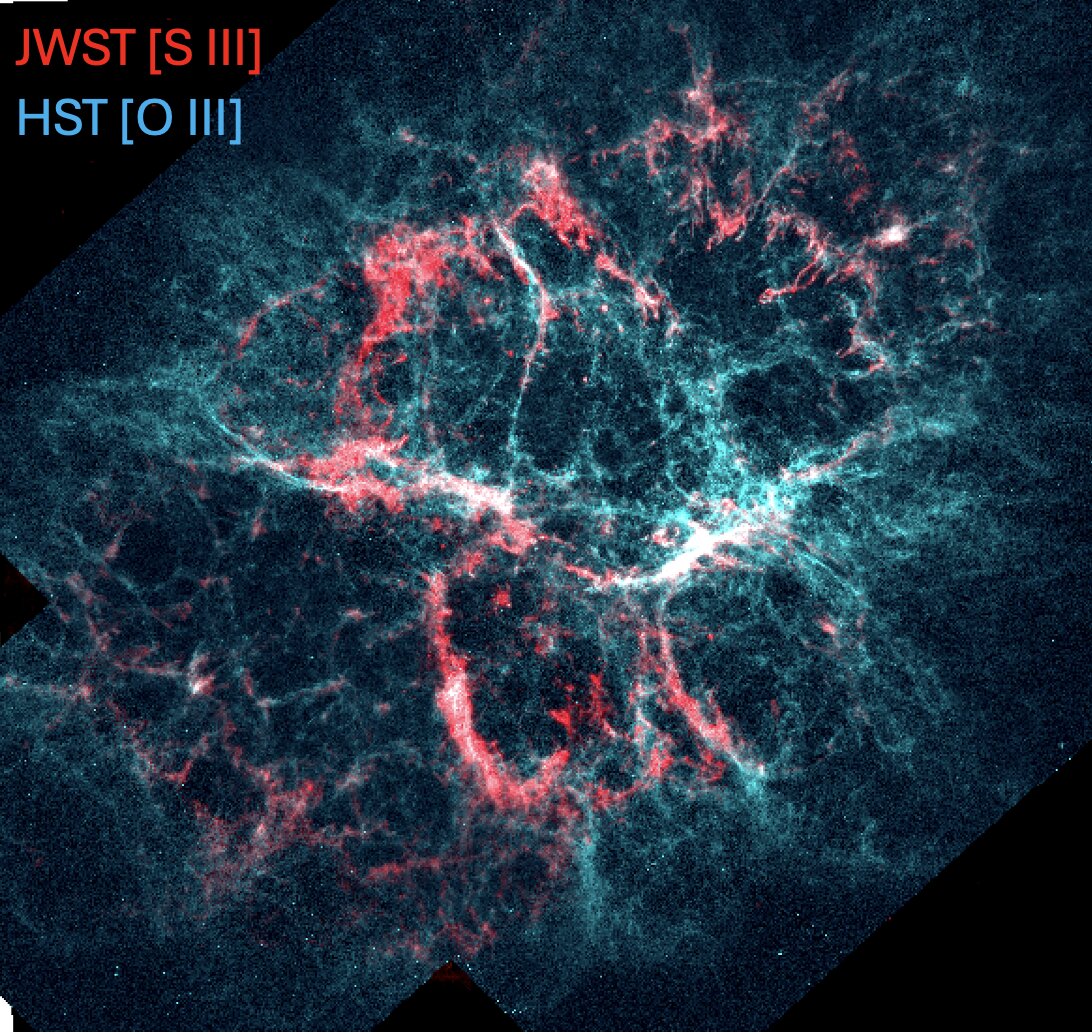}
\caption{This figure shows the higher ionization \oiii\ line in teal (ionization potential of $\rm O^{+}$ to $\rm O^{++}$ is 35.12 eV), and the JWST \siii\ line map in red (ionization potential of $\rm S^{+}$ to $\rm S^{++}$ is 23.34 eV).  Even though these lines are from different elements, many of the differences seen are due to varying ionization and thus density.  In particular, the \siii\ map is dominated by the cage filaments and the morphology of the emission is only slightly more extended than other emissions that trace the dense filaments.  The morphology of the \oiii\ is very different, sampling lower density/higher ionization gas, which is less clumpy and more extended.
\label{comp-s3o3}
}
\end{figure}

\begin{figure}
\center
\includegraphics[width=0.95\textwidth]{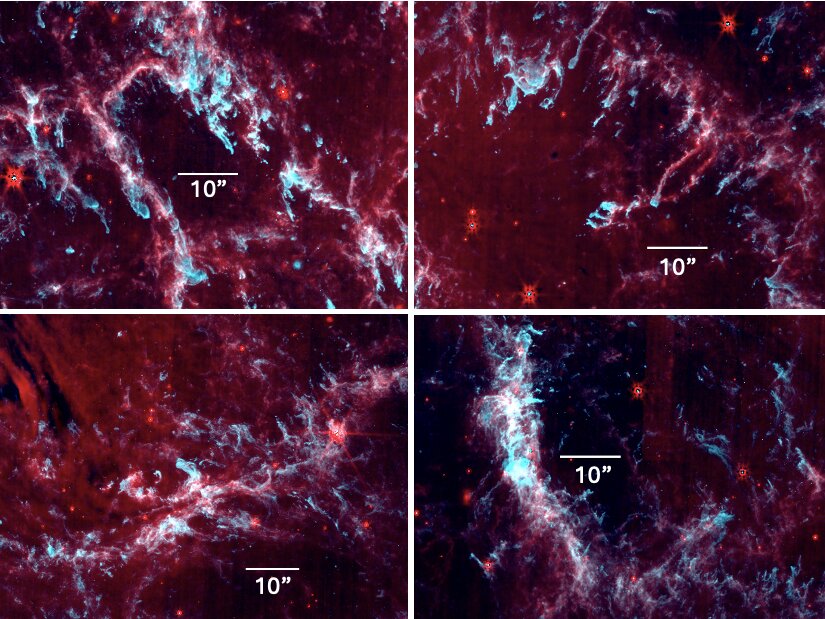}
\caption{
\label{fe2s2sections}
This figure shows four sub-regions of the JWST \feii\ (in red) and WFC3 \sii\ (teal), as shown in Fig.\ref{comp-fe2} (right panel) in the main paper. Since both of these ions have low ionization potentials, the differences seen here may be largely due to variable abundances.
}
\end{figure}

\clearpage

\end{document}